\documentclass[%
 aip,
 amsmath,amssymb,
 reprint,%
]{revtex4-1}

\usepackage{graphicx}
\usepackage{dcolumn}
\usepackage{bm}

\usepackage[utf8]{inputenc}
\usepackage[T1]{fontenc}
\usepackage{mathptmx}

\begin{document}

\preprint{AIP/123-QED}

\title[RNA in planetary dynamics]{Recurrence Network Analysis of Exoplanetary Observables}

\author{Tam\'as Kov\'acs}
 \email{tkovacs@general.elte.hu}
\affiliation{ 
Institute of Physics, E\"otv\"os University,
  P\'azm\'any P. s. 1A, Budapest 1117, Hungary
}%

\date{\today}

\begin{abstract}
Recent advancements of complex network representation among several disciplines motivated the investigation of exoplanetary dynamics by means of recurrence networks. We are able to recover different dynamical regimes by means of various network measures obtained from synthetic time series of a model planetary system. The framework of complex networks is also applied to real astronomical observations acquired by recent state-of-the-art surveys. The outcome of the analysis is consistent with earlier studies opening new directions to investigate planetary dynamics.
\end{abstract}

\maketitle

\begin{quotation}
The dynamical variety of known extrasolar planetary systems stimulates their stability analysis as an important characteristic of the system. The common procedure to perform such an investigation is to obtain the parameters of the best-fit planetary model and then integrate the equations of motion numerically. In this work we propose a method in order to describe system dynamics that is based on the observed uni-variate time series and the topology of complex network reconstructed from the signal itself. We show that the procedure is capable to distinguish the ordered and chaotic motion in a synthetic 2-planet system with a given level of significance. We also find that different kind of data sets, produced by the currently used detection methods, works well within the framework of the method. The analysis of real-world observations also provides results consistent with former studies. The technique is computationally very efficient since it does not require the phase space trajectories and, therefore, costly n-body simulation can be avoided. \textit{Thus this new strategy can be used as a complementary tool to extract the dynamical behavior of extrasolar planetary systems.} 
\end{quotation}

\section{\label{sec:intro}Introduction}
Dynamical stability is one of the most relevant physical characteristics (beside the bulk density, atmospheric composition, interiors of a planet, etc.) to describe extrasolar planetary systems. In order to study the dynamical evolution of a system the initial conditions of the numerical integration are essential. In planetary science this input includes the orbital elements for a given epoch. Nowadays the ground and space-based observations provide a large amount of extremely precise radial velocity (RV) and transit timing data. These information can be transformed by using comprehensive statistical methods \cite{Foreman-Mackey2013} into the desired initial conditions.

The majority of known exoplanetary systems harbour more than one planet resulting in complex non-Keplerian dynamics \cite{Fabrycky2010}. 
The possible instability of the system is an internal attribute of the deterministic dynamics when three or more celestial objects participate in the motion. To explore the chaotic nature one has to have the integrated trajectories and/or the structure of the relevant part of the phase space assisted by a chaos detection method (Lyapunov exponent, MEGNO ($\mathcal{M}$) -- Mean Exponential Growth of Nearby Orbits  \cite{Cincotta2000}, etc.). These requirements can be fulfilled by having the best-fit planetary model incorporating the orbital elements and the masses as it has already been shown \cite{Rivera2001,Armstrong2015,Batygin2015}. The results of these studies indicate the heterogeneity of exoplanetary dynamics. Interestingly, there are also candidates with chaotic properties \cite{Deck2012,Panichi2017} beside the larger number of stable resonant configurations.

We propose an alternative method to perform stability analysis of exoplanetary systems that requires only a scalar time series of the measurements, e.g. RV, transit timing variation (TTV) \footnote{As a consequence of the mutual gravitational interaction of the planets one might observe slight deviations in orbital periods called transit timing variation.}, or astrometric positions. The fundamental concept of Poincar\'e recurrences in closed Hamiltonian systems and the powerful techniques of nonlinear time series analysis \cite{Kantz2003} combined with complex network representation \cite{Donner2010} allow us to investigate the underlying dynamics without having the equations of motion. That is we can ignore the orbital elements as initial conditions and other system parameters from the entire analysis. Moreover, the procedure completely disregards the use of numerical n-body integration which is an essential and sometimes fairly time consuming part of the stability analysis. We present that this new scheme, first time applied in exoplanetary research, works well with signals acquired by the currently available observational techniques (see above).

\section{Method}
 A scalar time series carries relevant dynamical information if the measured physical quantity is coupled to other state space variables of the system. Based on this fact Takens' embedding theorem \cite{Takens1981} ensures that the phase space trajectory can be reconstructed from the measured data. The procedure is called time delay embedding and requires two parameters, the time lag ($\tau$) and the embedding dimension ($d$). There are standard techniques to find the appropriate values of these parameters, e.g. false nearest neighbours help to estimate the embedding dimension \cite{Kantz2003} while the first minimum of the mutual information function yields the time delay \cite{Semmlow}.

\begin{figure*}
\includegraphics[width=\textwidth,angle=0]{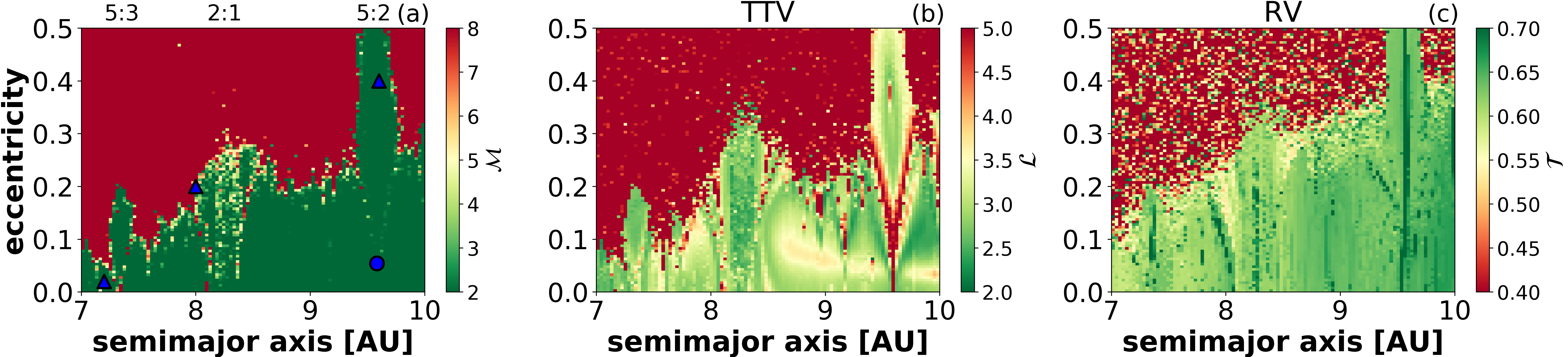}%
\caption{Stability and RN measures for the SJS system. (a) Stability map ($a_{\mathrm{Sat}},e_{\mathrm{Sat}}$) of two-planet system according to the indicator $\mathcal{M}.$ Semimajor axis is measured in astronomical units (au). The green area denotes the stable realm while for larger eccentricities the dynamics is chaotic (red). The dominant low-order mean motion resonances (MMR) are indicated at the top of the panel. Blue triangles are reference trajectories from different dynamical regimes for further analysis. The blue dot represents the Saturn's current location in the parameter plane. (b) and (c) Two RN measures $\mathcal{L}$, $\mathcal{T}$ are pictured as the results of RNA on the same grid as in (a) taking into account two observables TTV of Jupiter and RV of the Sun, respectively. The color bars illuminate the heat map values in each case. 
  \label{fig:1}}
\end{figure*}

Once the reconstructed trajectory is ready to use, one can search for Poincar\'e recurrences, the events when the trajectory $\mathbf{x}$ comes back at time $t_j$ into an $\epsilon$-ball around its earlier position at time $t_i$ ($t_i<t_j$). The idea of recurrence plot (RP) \cite{Eckmann1987} is a 2D binary matrix representation of these events. More precisely, the matrix element $R_{i,j}(\epsilon)=1$ if $||\mathbf{x_j}-\mathbf{x_i}||<\epsilon,$ and 0 otherwise. The $||.||$ can denote various norms such as Euclidean, Maximum, Manhattan. In this work the Maximum norm is used through the whole analysis. The $\mathbf{R}$ matrix is symmetric by its nature. First Refs.~\cite{Zbilut1992,Webber1994} \footnote{The visual interpretation of the matrix $\mathbf{R}$ can achieved by plotting a dot when the matrix element is 1 and leave it empty otherwise.} proposed the method of recurrence quantification analysis (RQA) which identifies various measures based on the diagonal and horizontal/vertical texture of an RP. The advantage of RQA in dynamical analysis has been demonstrated in many applications from climatology through neuroscience to astrophysics. The idea of recurrence networks (RN) \cite{Donner2010,Donner2011} revolutionized the RP representation of dynamical systems using the fact that an RP can be thought of as the adjacency matrix of a complex network embedded in phase space. Quantitatively
\begin{equation}
A_{i,j}=A_{i,j}(\epsilon)=R_{i,j}(\epsilon)-\delta_{ij},
\label{eq:adjmat}
\end{equation}
where $A_{i,j}$ is the adjacency matrix, $R_{i,j}$ the recurrence plot, and $\delta_{ij}$ the Kronecker delta. The role of $\delta_{ij}$ is to exclude the loops. The nodes of these graphs are the state space vectors that are connected by edges if their proximity is smaller than a threshold $\epsilon.$ The main advantage of the complex network framework is that the temporal correlations can be avoided since the dynamical feature of the underlying system is preserved via its topology, thus, the explicit time ordering is not relevant. The standard quantities known in graph theory (such as degree centrality, average path length, transitivity, etc.) are also applicable to RNs and they are capable to describe different dynamical regimes and dynamical transitions \cite{Zou2019}. This fact will be utilized in the following.

\section{Results}
\subsection{The model system}
 First a well-defined model is investigated in order to present the reliability of the method. Consider the Sun-Jupiter-Saturn (SJS) two-planet system in our own Solar System with their actual orbital elements. One can explore, for example with the \texttt{REBOUND} package \cite{Rein2015}, the stability of the system by varying, say, Saturn's initial semimajor axis ($a_{\mathrm{Sat}}$) and eccentricity ($e_{\mathrm{Sat}}$) in a grid and quantify the dynamics. Figure~\ref{fig:1}(a) shows the result of dynamical stability of the system in the ($a_{\mathrm{Sat}},e_{\mathrm{Sat}}$) initial condition plane for 100x100 different trajectories characterized by the chaos indicator MEGNO ($\mathcal{M}$). The values of $\mathcal{M}\approx 2$ indicate stable motion while larger values (in this case truncated at 8) represent chaotic dynamics. One can observe that for higher eccentricities chaotic nature dominates the phase space. Nevertheless, there are tight green lobes penetrating into the red area. These formations are the MMRs between the two planets. The most remarkable resonance, 5:2, around $a_{\mathrm{Sat}}\approx 9.6$ extends up to very high eccentricities. The stability map of the SJS system is based on numerical integration that cover ca. 1000 orbital periods of Jupiter for each initial condition pairs. This astronomically fairly short time frame has been chosen in order to mimic real observations. The compact extrasolar planets, for instance the TRAPPIST-1 \cite{Gillon2017}, Kepler-412 \cite{Deleuil2014}, make thousands of revolutions on human time scales producing time series with desired length to be analyzed: $\sim$1000 periods of the innermost planet.

\begin{table*}
\caption{\label{tab:table1}The embedding parameters of various time series in SJS system used to construct RNs. Initial conditions in Fig~\ref{fig:1}(a) are considered with two different length, the original 950 data point signals and longer one containing 3500 measurements. The time delay parameters $d,$ $\tau,$ and $\epsilon$ are the embedding dimension, time lag, and threshold, respectively. The threshold in each case is obtained from recurrence matrix with fixed $RR=0.1.$}
\begin{ruledtabular}
\begin{tabular}{lcdddddd}
 &&\multicolumn{2}{c}{Regular}&\multicolumn{2}{c}{Chaotic}&\multicolumn{2}{c}{Resonant}\\
 Data&&950&3500&950&3500&950&3500\\ \hline
 RV-Syn\footnote{Synthetic data are obtained directly from numerical integration.}&$d$ &4&6&10&9&5&5\\
 &$\tau$ &2&2&3&2&3&2\\
 &$\epsilon$ &2.1e-4&1.84e-4&3.3e-4&2.5e-4&1.5e-4&2.1e-4\\
 RV-Spl\footnote{Spline interpolation has been made on synthetic (Syn) time series after adding noise and removing some data points as described in the text.}&$d$ &10&11&14&13&11&11\\
 &$\tau$ &3&5&3&2&3&3\\
 &$\epsilon$ &5.65e-6&4.8e-6&5.72e-6&5.0e-6&5.7e-6&4.6e-6\\
 TTV-Syn&$d$ &8&9&18&17&7&9\\
 &$\tau$ &4&4&5&4&10&19\\
 &$\epsilon$ &251.8&292.8&4922.05&6814.92&340.35&1019.59\\
 TTV-Spl&$d$ &9&12&15&16&13&17\\
 &$\tau$ &3&3&12&14&11&16\\
 &$\epsilon$ &34.05&53.51&634.87&987.25&47.38&374.74\\
\end{tabular}
\end{ruledtabular}
\end{table*}

\begin{figure}
\includegraphics[width=\columnwidth,angle=0]{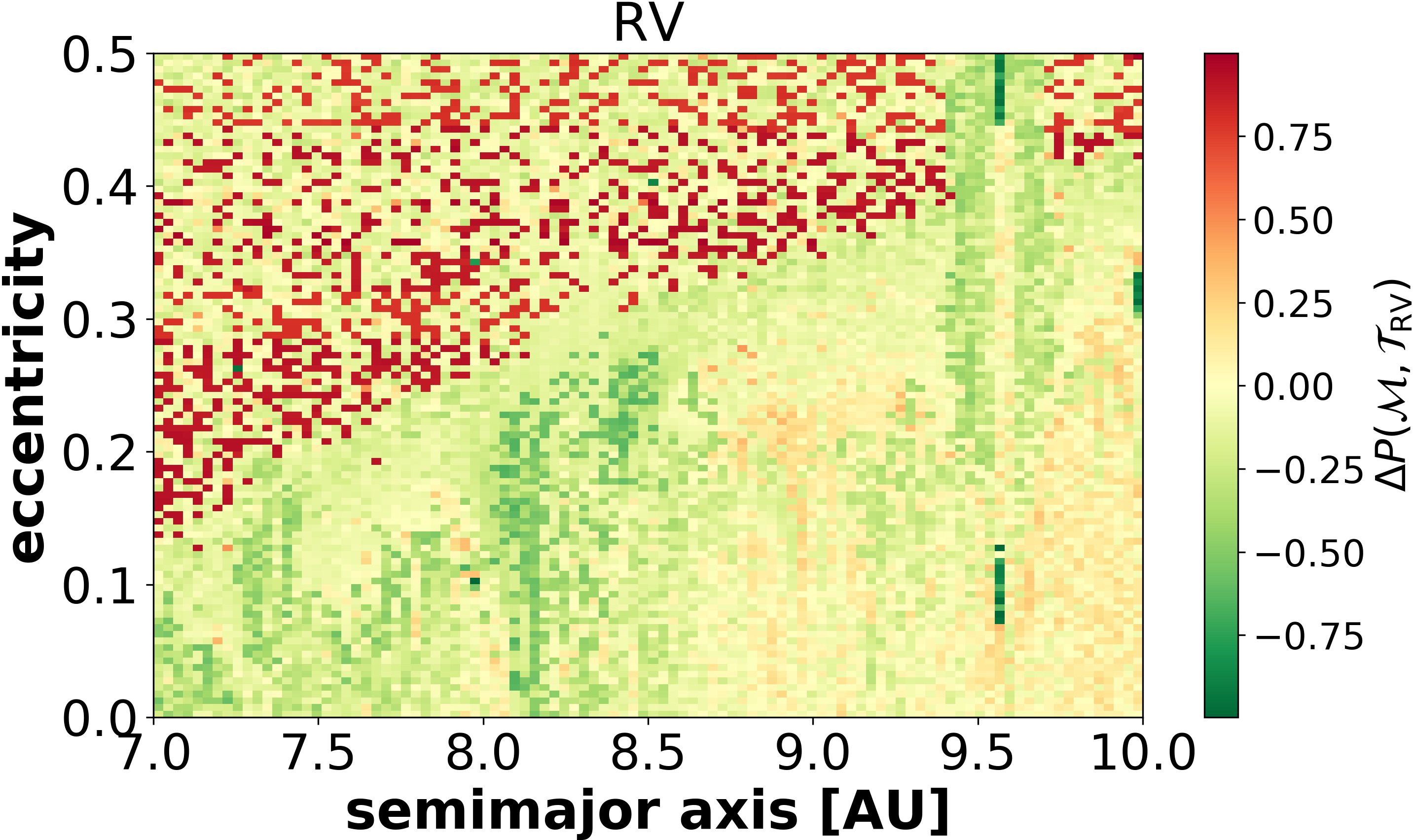}%
\caption{CDF differences between the chaos indicator $\mathcal{M}$ and RN measure $\mathcal{T}_{\mathrm{RV}}$ in the parameter plane ($a_{\mathrm{Sat}},e_{\mathrm{Sat}}$).
  \label{fig:rev1}}
\end{figure}

In order to apply recurrence network analysis (RNA) to SJS system, synthetic scalar time series were acquired from the full phase space trajectories in the following way. Radial velocity data is the x-component of the Sun's velocity vector while transits are taken when the planet crosses the positive x-axis imitating the position of a real occultation. In case of the RV signals the trajectories were integrated until 1050 Jupiter orbits and 950 data points were sampled in order to set the sampling frequency larger than the mean motion of the Jupiter. For transits exactly 950 events have been taken. This also implies that the trajectories examined via transits have somewhat different length, namely, 950 Jupiter transits might take shorter or longer time in different dynamical regimes. 

\begin{figure*}
\includegraphics[width=\textwidth,angle=0]{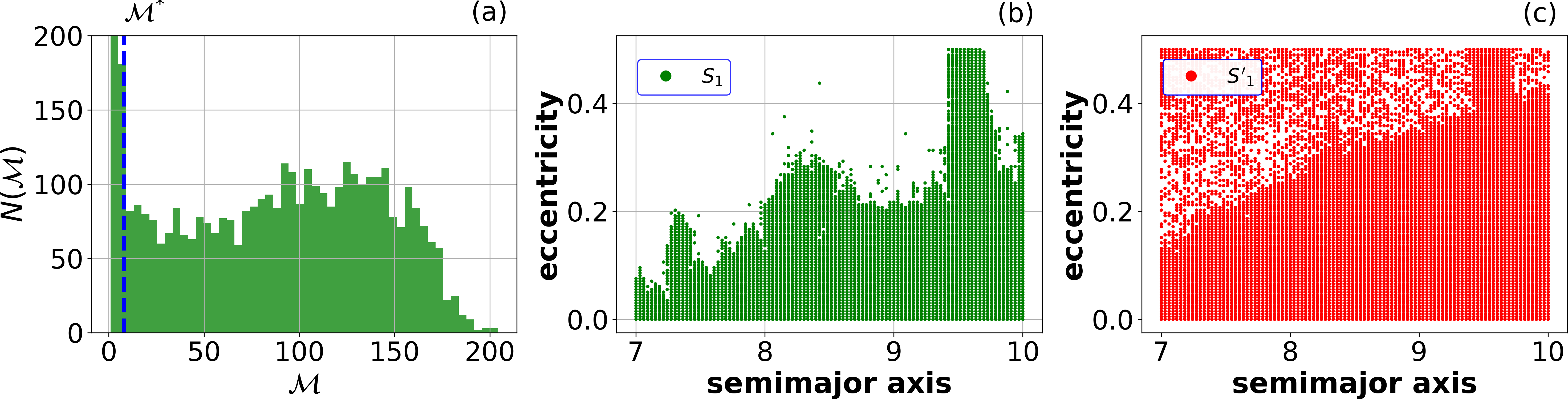}%
\caption{(a) Distribution of chaos indicator MEGNO using the 100x100 grid of ($a,e$) parameter plane. (b) and (c) Two particular subsets of $S_1$ and $S'_1$ obtained from the same quantile defined by $\mathcal{M}^*=8.$
  \label{fig:rev2}}
\end{figure*}

\begin{figure}
\includegraphics[width=\columnwidth,angle=0]{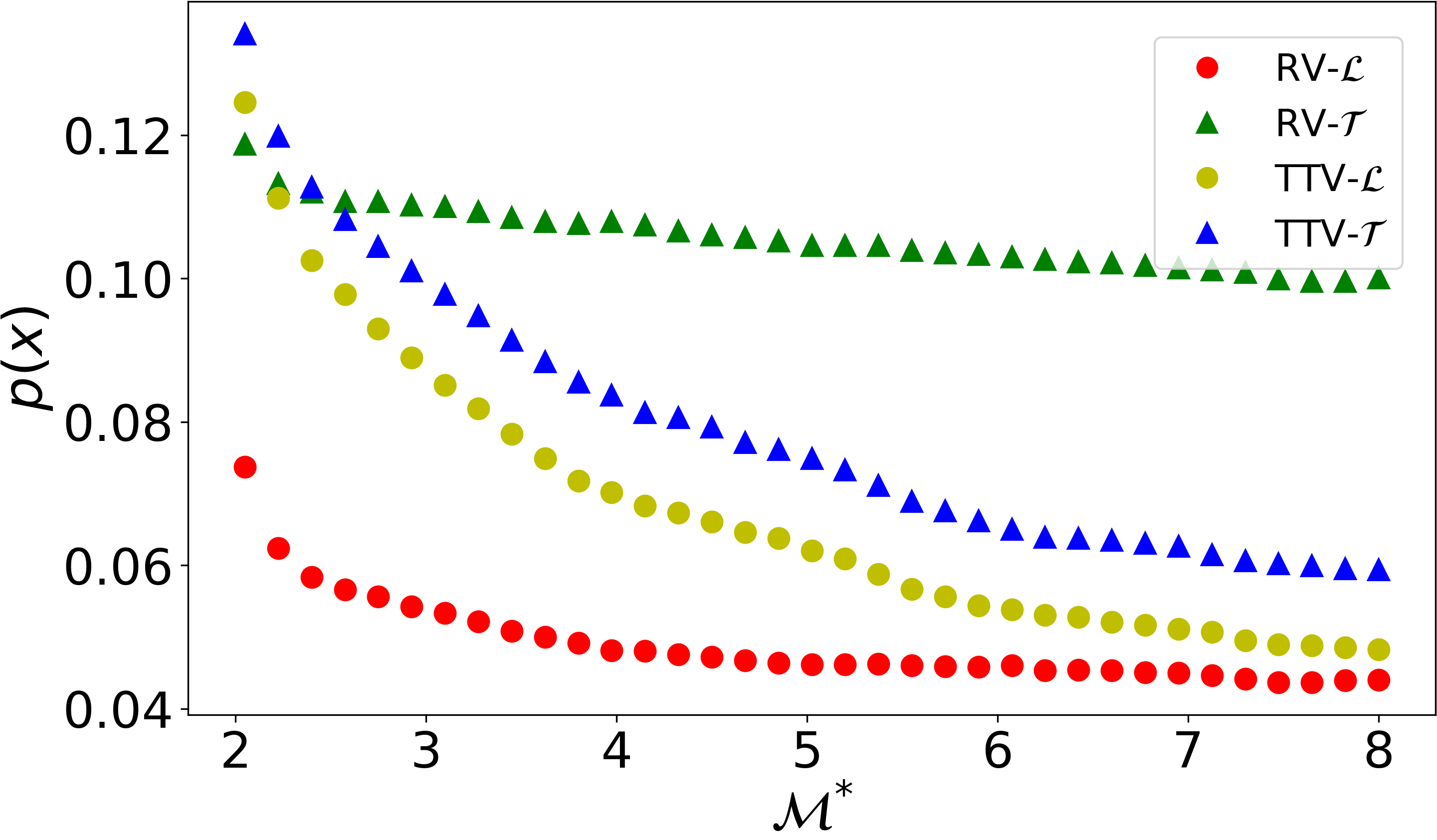}%
\caption{The relative frequency of false detection of various characteristics vs. $\mathcal{M}^*$ based on the same quantile. 
  \label{fig:rev3}}
\end{figure}

Figure~\ref{fig:1}(b) and (c) depict the network measures average path length ($\mathcal{L}$) for TTV signals and transitivity ($\mathcal{T}$) for RV data, respectively, in the ($a_{\mathrm{Sat}}, e_{\mathrm{Sat}}$) parameter plane. The phase space reconstruction and the network analysis has been done by the \texttt{TISEAN} \cite{Hegger1999} and \texttt{PYUNICORN} \cite{Donges2015} packages, respectively. The similarity is evident between the panels (a) and (b), (c). In panel (b) for TTV signals the larger the $\mathcal{L}$, the more chaotic the dynamics. This result is in good agreement with general findings in discrete Hamiltonian systems like the standard map\cite{Zou2016}. The RNA of RV time series produces similar results for the measure $\mathcal{T}$, panel (c). In this case transitivity closer to 1 yields weaker instability according to the general picture that regular trajectories show smaller divergence in phase space. This fact results in a more coherent topology of the network that produces higher transitivity. One can also notice that the stability map of TTVs is more sensitive to the edge of chaos inducing more details between the green and red domains. This is also important because weak chaos, also known as stickiness \cite{Contopoulos2002} in Hamiltonian dynamics, emerges just at the edge of stability islands. The stickiness effect was also identified by using complex network measures in low dimensional conservative systems \cite{Zou2016}. The method of RNA applied to single variable time series is, thus, able to reconstruct the stability map what is obtained from numerical integration and phase space trajectories. Moreover, it is also remarkable that achieving this outcome the length of the measured data must not be longer than that of the phase space trajectory.

Table~\ref{tab:table1} contains the embedding parameters for the reference trajectories introduced in Figure~\ref{fig:1}. In addition, time delay embedding has been done for longer time series containing 3500 data points in all three cases. This analysis verifies that 950 RV and TTV measurements provide a reliable RNA. In order to demonstrate that RN measures ($\mathcal{L},\mathcal{T}$) are suitable to distinguish regular and chaotic orbits, further statistical analyses are performed.

First, a point-wise difference based on the cumulative distribution functions (CDF) of $\mathcal{M}$ and ($\mathcal{L},\,\mathcal{T}$) in the ($a_{\mathrm{Sat}},e_{\mathrm{Sat}}$) parameter plane is designed \cite{Zou2010}
\begin{equation}
    \Delta P(\mathcal{M},x)=P(\mathcal{M})-P(x),
\end{equation}
where $P(x),$ $x\in \{\mathcal{L}_{\mathrm{RV}},\mathcal{T}_{\mathrm{RV}},\mathcal{L}_{\mathrm{TTV}},\mathcal{T}_{\mathrm{TTV}}\},$ is the corresponding value of CPD at each combinations of 10000 ($a_{\mathrm{Sat}},e_{\mathrm{Sat}}$) pairs. An example of CDF differences, $\Delta P(\mathcal{M},\mathcal{T}),$ for RV data is depicted in Figure~\ref{fig:rev1}. The difference is close to zero almost in the entire parameter plane. Remarkable deviation from zero can be observed in chaotic region.

Another more sophisticated quantitative comparison of RN measures can be obtained as follows. Let us define two disjoint subsets of MEGNO distribution defined by a critical value of $\mathcal{M}^{*}$
\begin{equation}
    \begin{split}
        S_1(\mathcal{M}^{*})&:=\{(a,e)|\mathcal{M}(a,e)\leq\mathcal{M}^{*}\},\\
        S_2(\mathcal{M}^{*})&:=\{(a,e)|\mathcal{M}(a,e)>\mathcal{M}^{*}\},\\
    \end{split}
\end{equation}
with group size $n_1$ and $n_2=n-n_1$ ($n=10000$), respectively, see Figure~\ref{fig:rev2}(a). Then, take the $\alpha$-quantile $Q_{\alpha}(\mathcal{M})$ with $\alpha=n_{1}/n$ of the distribution $\mathcal{M}$ for a given $\mathcal{M}^{*}$ and construct the same division for the corresponding ($a_{\mathrm{Sat}},e_{\mathrm{Sat}}$) parameters based on various RN measures
\begin{equation}
    \begin{split}
        S'_1(Q_{\alpha}(x))&:=\{(a,e)|x(a,e)\leq Q_{\alpha}(x)\},\\
        S'_2(Q_{\alpha}(x))&:=\{(a,e)|x(a,e)>Q_{\alpha}(x)\},\\
    \end{split}
\end{equation}
where $x$ again incorporates certain network characteristics. This analysis allows one to compare the difference of two distributions based on distinct measures. Figure~\ref{fig:rev2}(b) and (c) depict one particular example of two subsets of ($a_{\mathrm{Sat}},e_{\mathrm{Sat}}$) pairs associated to $S_1(\mathcal{M}^{*}=8)$ and $S'_1(Q_{\alpha}(\mathcal{T}_{\mathrm{RV}})),$ $\alpha\approx 0.57.$

The relative frequency $p$ of those parameter pairs that do not belong to the same group based on the two different measures indicate the false detection of dynamical nature. Slightly varying the chaos indicator close to the border of regular and chaotic feature, $2\leq\mathcal{M}^*\leq 8,$ the relative frequency of ''grouping errors'' can be quantified. Figure~\ref{fig:rev3} collects the frequencies of the average path length and transitivity obtained from RV and TTV signals representing that the classification error remains under 10\% in all cases.
 
Although, the RN measures give qualitatively reasonable success for either observables, it is clear that their values depend on the choice of the threshold $\epsilon$. To avoid this weakness two possibilities are known. First, one can derive dynamical invariants (e.g. R\'enyi entropy, correlation dimension, maximal Lyapunov exponent) from recurrence plots. However, to get reliable feedback about the dynamics by these invariants one has to have much longer time series. And this is against our will, to make stability analysis based on real observables that contain only several hundreds/thousands of data points. The second option is to perform hypothesis tests with surrogate time series \cite{Schreiber2000,Luo2005}. Since in planetary dynamics the signals show predominantly quasi-periodic variations, the null hypothesis in surrogation method we want to test should be that the original data comes from quasi-periodic process. Pseudo-periodic twin surrogates (PPTS) \cite{Carrion2016} combine the power of Pseudo-periodic \cite{Small2001} and Twin surrogate \cite{Thiel2006} methods and fit perfectly to test quasi-periodicity in planetary dynamics.

\begin{figure*}
\includegraphics[width=\textwidth,angle=0]{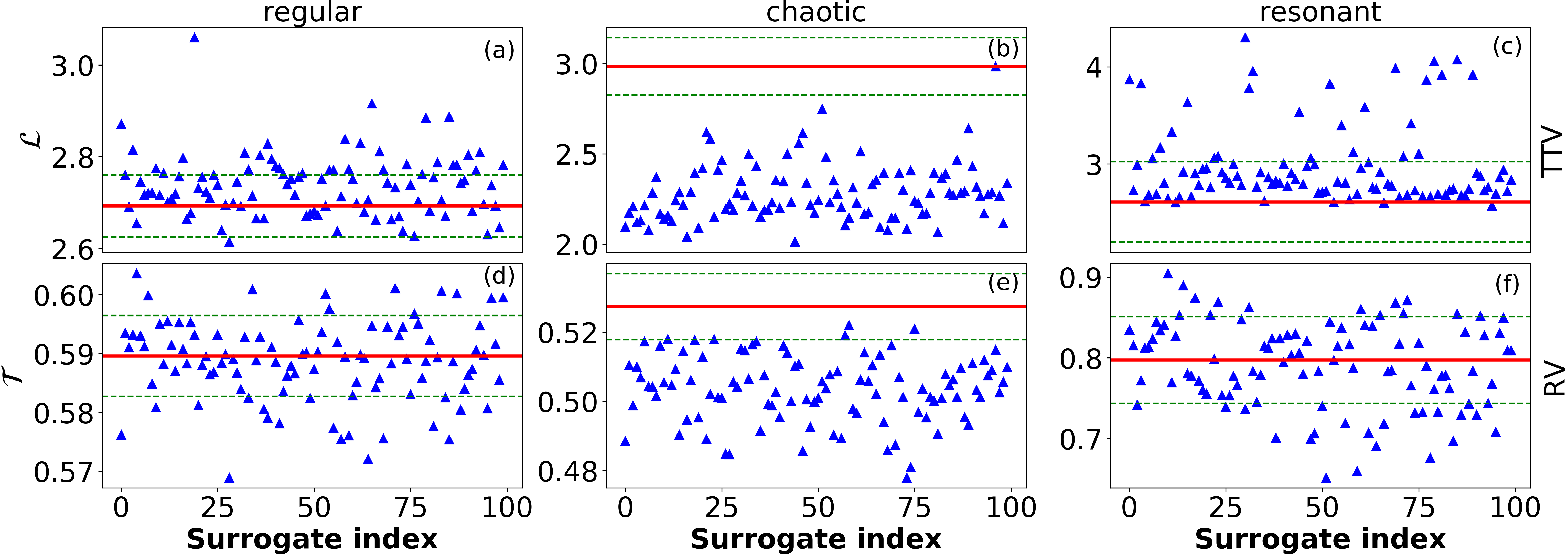}%
\caption{Outcome of hypothesis tests in three different dynamical regimes. Panels (a)-(c) show the results for $\mathcal{L}$ based on TTV signals. Panels (d)-(f) present the measures $\mathcal{T}$ in case of RV data sets. The rank-based statistics involves noisy and irregularly sampled reference trajectories (red solid line) and 100 PPT surrogates (blue triangles). The green dashed lines mark the $\pm$1 standard deviation of RN measures determined from the surrogate time series.\label{fig:2}}
\end{figure*}

\begin{figure*}
\includegraphics[width=\textwidth,angle=0]{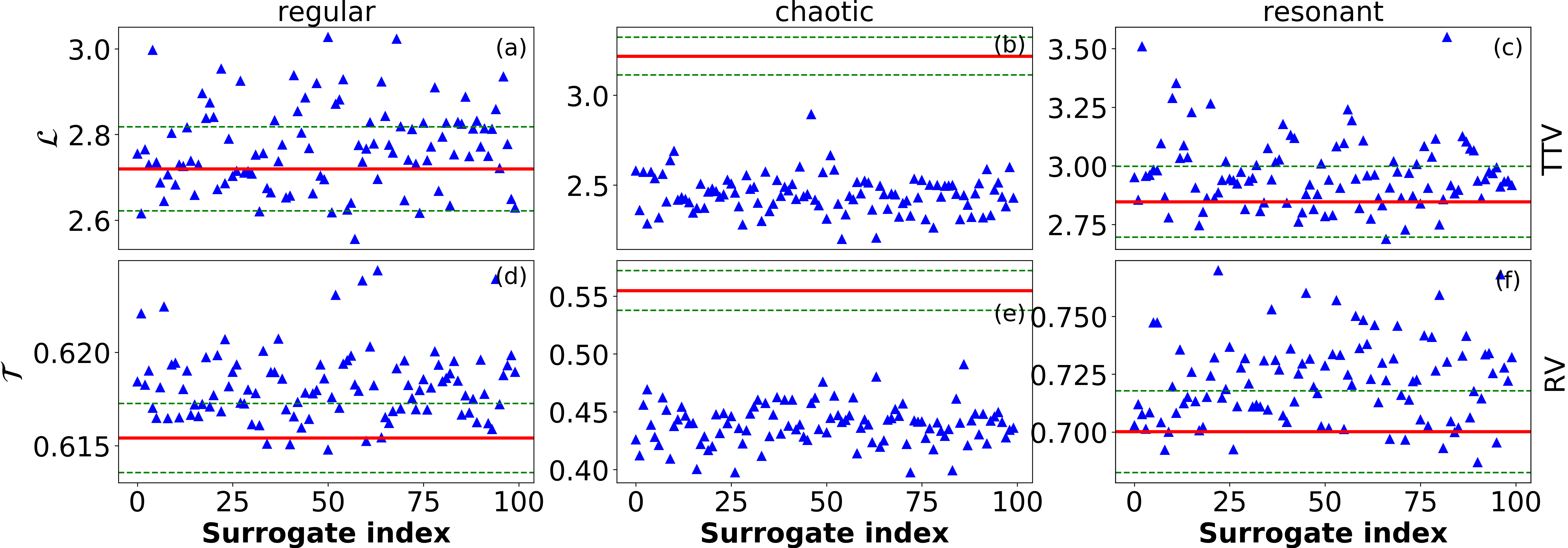}%
\caption{The same as in Figure~\ref{fig:2} for synthetic RV and TTV time series, top and bottom row, respectively. \label{fig:7}}
\end{figure*}

More concretely, time series generated by the PPTS method provide different RP matrices for periodic/quasi-periodic orbits and the chaotic ones preserving the phase space structure. This ensures that these surrogates are relevant to test the null hypothesis that the original signal comes from quasi-periodic process. The question, however, still remains open: What are the dynamical properties of the underlying system in case of rejecting the null hypothesis. In view of the fact that planetary dynamics is deterministic, high dimensional and definitely nonlinear and by declining the null hypothesis of PPTS, meaning that the motion is not quasi-periodicity, the dynamics should be chaotic according to certain confidence level.

The hypothesis tests in Figure~\ref{fig:2} comprise also the robustness of the RNA against measurement noise and missing data points. In order to reproduce a realistic astronomical observation, first Gaussian white noise with zero mean and unit variance is superimposed onto the synthetic time series in the SJS system with Signal-to-Noise ratio $\approx$6-7. Then, in a systematic way a random point is chosen along with an interval around this point which with random length is deleted from the data. This procedure is repeated until the desired amount of information (15\%) is missing from the original signal. Based on \cite{Lekscha2018}, spline interpolation is performed on the unevenly spaced data before the time delay embedding and RNA is applied.

To obtain a 1\% level of significance for a one-sided test 99 surrogate time series must be generated \cite{Kantz2003} and then the RN measures calculated from the original time series has to be compared to those acquired from the surrogates. Usually we do not know whether the distribution of RN measures is Gaussian, therefore, a rank-based statistics is preferred instead of a normal. A rank-based statistics of the RNA specifies either to keep or reject the null hypothesis. Figure~\ref{fig:2} illustrates the hypothesis tests for three different kind of dynamics marked by blue triangles in Fig.~\ref{fig:1}(a), and two observables (RV, TTV) in the SJS system. Let us concentrate on the first column, panels (a) and (d). The analysis here is devoted to show that the leftmost blue triangle ($a_{\mathrm{Sat}}, e_{\mathrm{Sat}}$)=(7.2,0.02) corresponds to regular quasi-periodic motion. The red solid lines represent the RN measures $\mathcal{L}$ and $\mathcal{T}$ gained from the synthetic time series while the blue triangles symbolize the same RN measures based on the generated 100 PPTSs. Both plots indicate that $\mathcal{L}$ and $\mathcal{T}$ associated to the original time series fall into the ensemble of blue triangles. Thus, one can keep the null hypothesis, i.e. the original signal comes from quasi-periodic dynamics.

Due to strong gravitational perturbation taking place primarily at large eccentricities the system is destroyed, i.e. one of the planets escapes. Therefore, an initial condition close to the border of the regular part  has been considered where the motion is chaotic while bounded for the integration time, ($a_{\mathrm{Sat}}, e_{\mathrm{Sat}}$)=(8.0,0.2). Executing the RNA one finds the results portrayed in panels (b) and (e). The scheme is the same as before, in turn, it can be clearly seen that the red lines are located outside the blue triangle zoo suggesting that the null hypothesis can be rejected. That is, the original time series was generated by chaotic dynamics. For the RNA we pick up a third, resonant, initial condition in the stability map at position ($a_{\mathrm{Sat}}, e_{\mathrm{Sat}}$)=(9.6,0.4). The reassuring results of the hypothesis tests are summarized in panels (c) and (f). 

In order to verify that the spline interpolation does not cause any artificial effect during time delay embedding, the same analysis has been made on synthetic, i.e. noiseless uniformly sampled, signals as well. Figure~\ref{fig:7} presents the outcome of hypothesis tests that perfectly match those based on noisy and scanty analysis in Figure~\ref{fig:2}.

\subsection{Real-world measurements}
\begin{figure}
 \includegraphics[width=\columnwidth,angle=0]{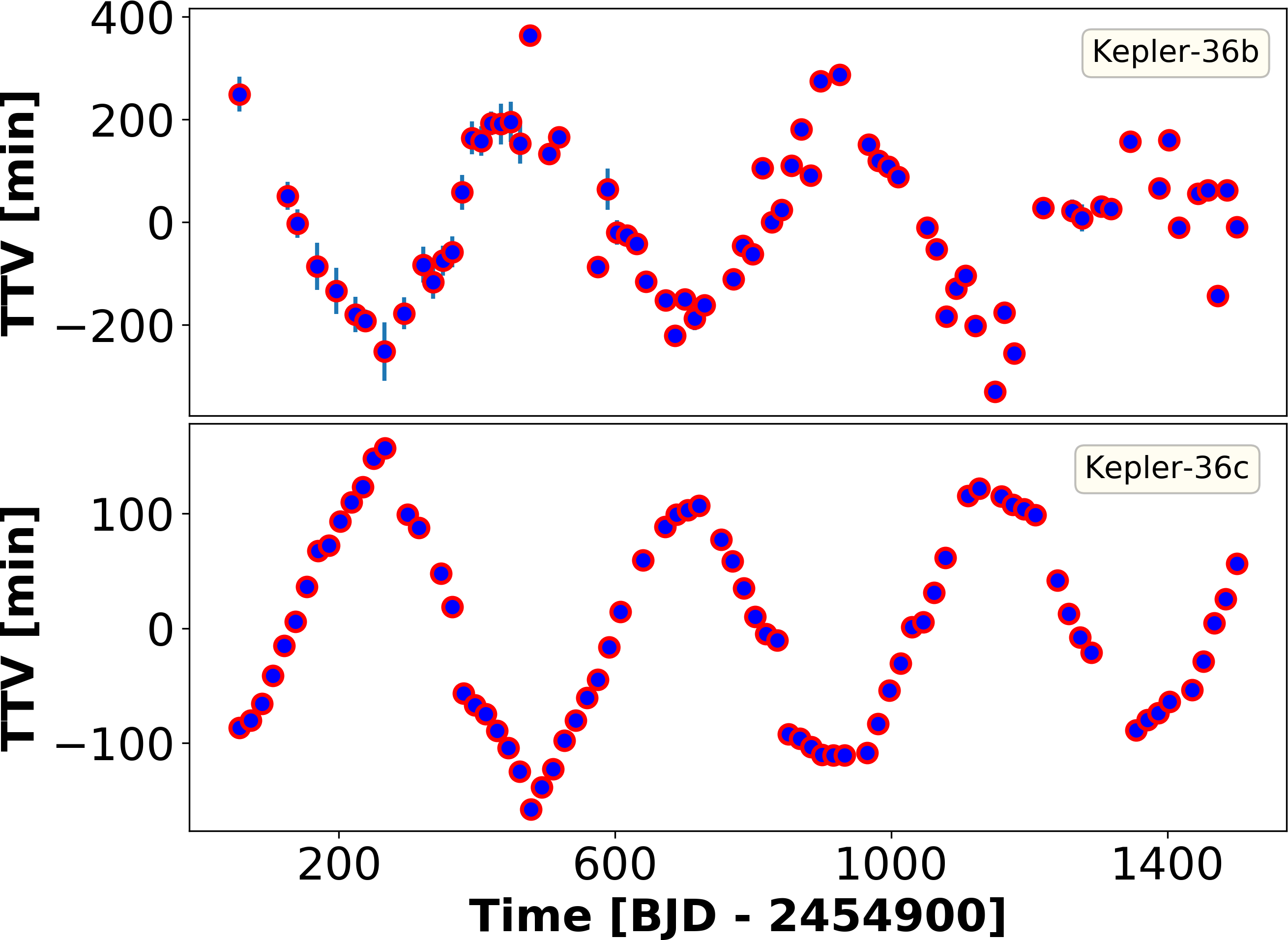}%
 \caption{Transit timing variation of Kepler-36b and c. The two signals are in anti-phase as is expected from the dynamical considerations. Upper panel: Kepler-36b the inner planet with smaller mass ($0.0135 M_{\mathrm{Jupiter}}$) and consequently larger TTV amplitude, 72 measurements during 103 epochs yield ~30\% missing data points. Lower panel: Kepler-36c the more massive outer planet ($0.0254 M_{\mathrm{Jupiter}}$) with better coverage, 77 observations out of 89 epochs, ~15\% missing data.
\label{fig:6}}
 \end{figure}

Recently the number of known extrasolar planetary systems outstandingly increased due to the cutting-edge technology. Extremely precise light curves with transit timing measurements might shed light on the dynamical diversity of the observed systems as shown above. The catalog of the full-cadence data set of the Kepler mission \cite{Holczer2016} contains a large number of dynamically interesting systems especially from the TTV point of view. Some of them have been extensively studied by means of stability \cite{Deck2012,Panichi2017}. One of these systems is Kepler-36 \cite{Carter2012} wherein two planets (Kepler-36b and c) orbiting the central Sun-like star nearly in 7:6 MMR. The orbital separation of the two ''Super-Earths'' is fairly small, 0.013 au. Due to the tightly packed configuration, the mutual gravitational interaction causes large TTV, see Figure~\ref{fig:6}. The irregular dynamics of the system was proposed first by Deck et al. \cite{Deck2012}, further analysis revealed \cite{Panichi2017} that the system might be close to the border of the 7:6 MMR and the dynamics is governed by the stickiness of the resonance.

\begin{figure}
\includegraphics[width=\columnwidth,angle=0]{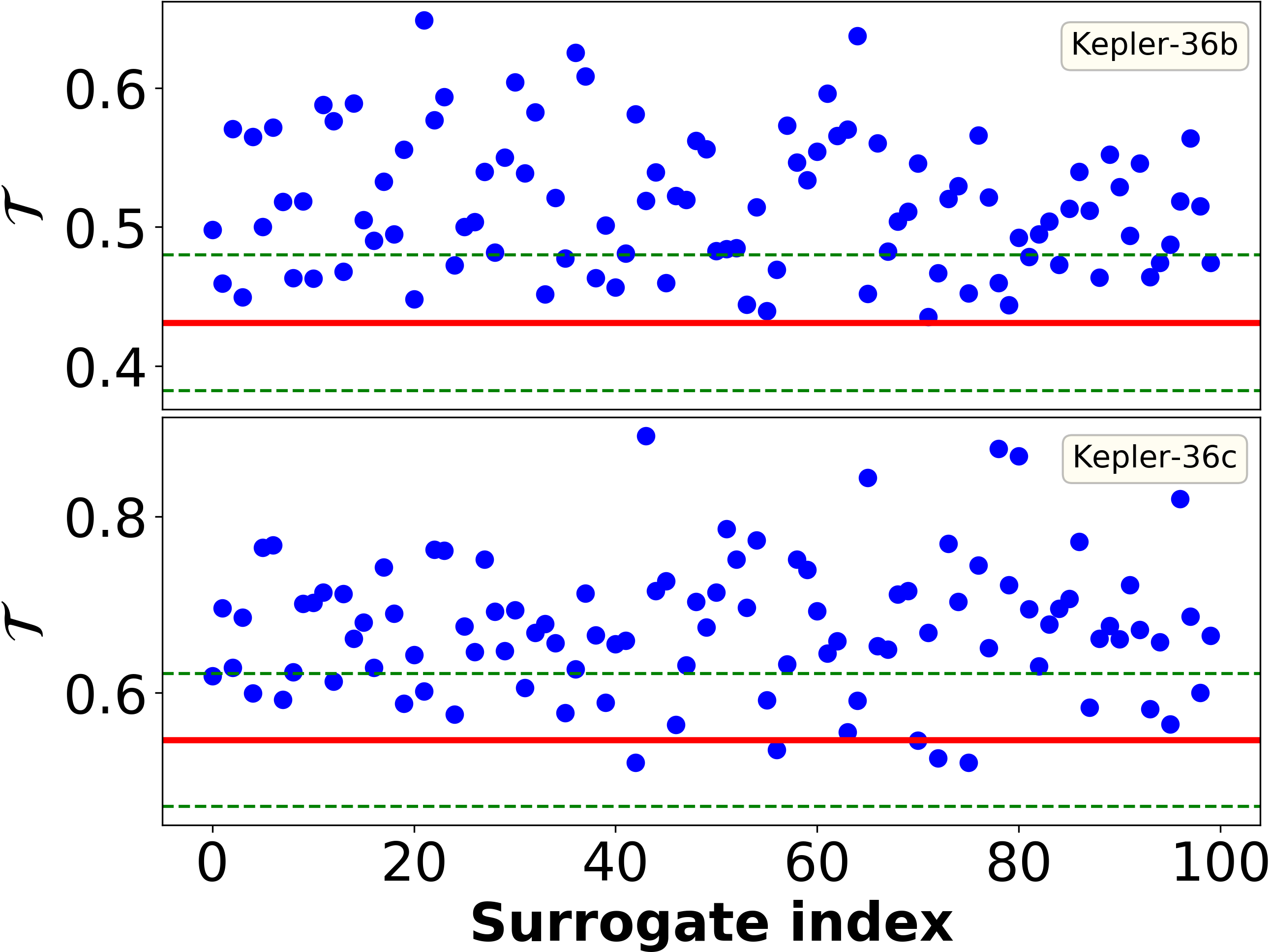}%
\caption{Hypothesis tests including RN measures $\mathcal{T}$ for both planets in Kepler-36 systems. The embedding parameters are $d=6,$ $\tau=4,$ and $\epsilon=165.$ The red solid lines represent the transitivity values obtained from the original signals, $\mathcal{T}_{b}=0.430,\,\mathcal{T}_{c}=0.546.$ The 100 surrogates, blue points, permit one-sided 99\% level of significance. The null hypothesis can be rejected in case of Kepler-36b as the rank based statistics suggests.\label{fig:3}}
\end{figure}

\begin{figure}
\includegraphics[width=\columnwidth,angle=0]{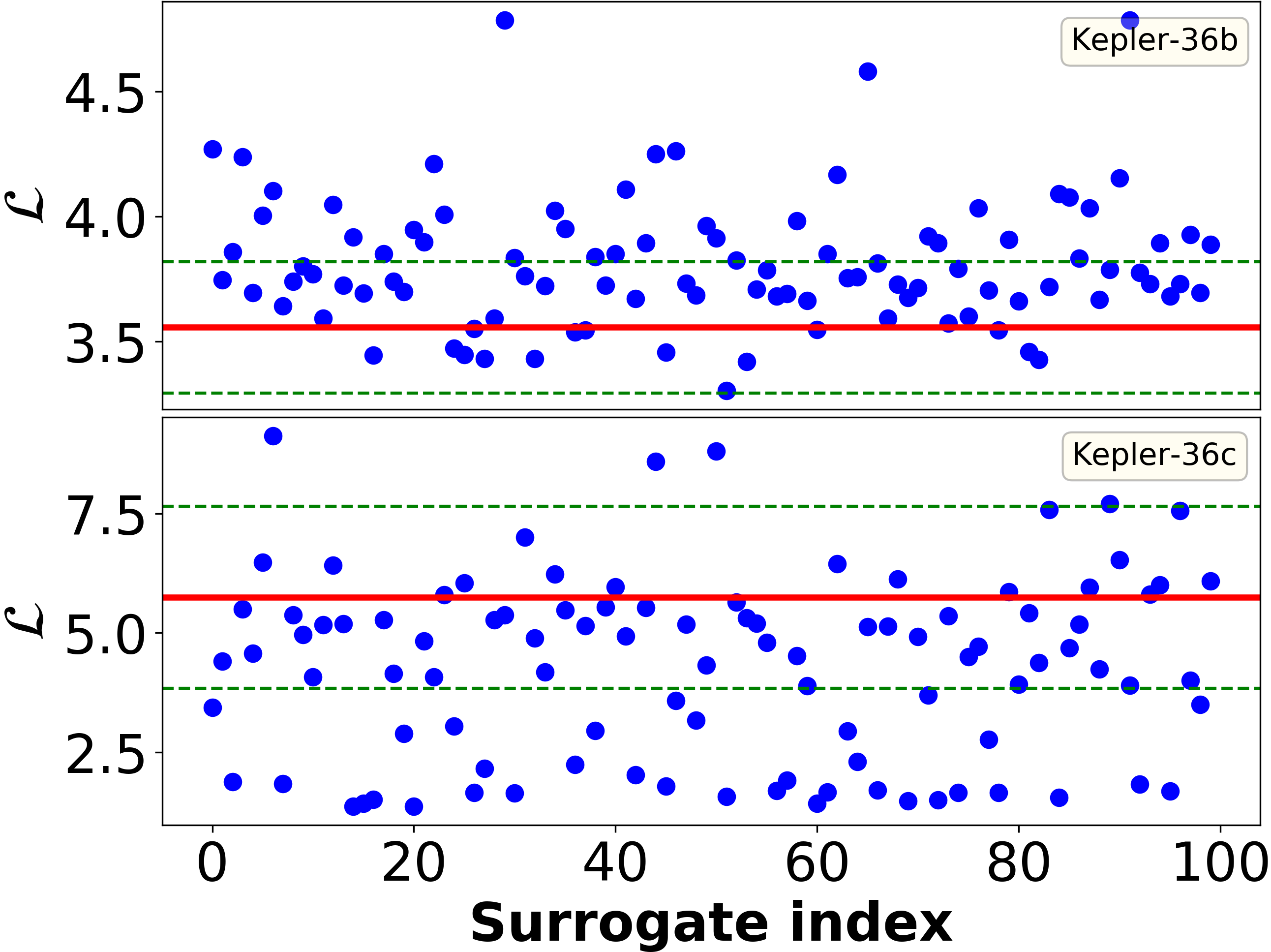}%
\caption{The same as in Figure~\ref{fig:3} The embedding parameters are $d=6,$ $\tau=5,$ and $\epsilon=52.90.$ The average path length values obtained from the original signals, $\mathcal{L}_{b}=3.557,\,\mathcal{L}_{c}=5.749.$ According to null hypothesis the dynamics is quasi-periodic.}\label{fig:4}
\end{figure}

Recurrence network analysis has been carried out for the TTV data of Kepler36b and c. It should also be noted that the data points in Kepler-36 time series are less with a factor of 8-10 than those in the SJS system. That is, the stability analysis is based on $\sim$100 orbital periods of the planets. Moreover, the number of missing transit points is somewhat larger than in the synthetic SJS, roughly 13\% for inner planet and 25\% for the outer one. The 99\% significance level hypothesis test involving RN measures $\mathcal{T}$ and $\mathcal{L}$ is presented in Figure~\ref{fig:3} and \ref{fig:4}. Results depicted in Fig.~\ref{fig:3} lower panel show that in the case of the more massive outer planet the dynamics is regular. In contrast, for the inner planet, upper panel, the null hypothesis can be rejected in accordance with the rank-based statistics described above. The outcome of the RNA implies that the system's behaviour is irregular. However, average path length for the same system in Fig.~\ref{fig:4} stipulates regular dynamics for both planets. 
Comparing the results with Figure~\ref{fig:1}, the statistical description of Kepler-36 fits to the picture of stable chaos appearing at the border of the resonances. Moreover, it also supports the view published in \cite{Deck2012,Panichi2017} based on different methods of dynamical analysis. 

\section{Conclusion}
We conclude that the RNA method is suitable to reproduce the dynamical behaviour of a planetary system with reliable significance. This result is based on the fact that the topology of recurrence networks preserves the underlying dynamics of the system producing the time series under study. Thus, the RN measures, $\mathcal{L}$ and $\mathcal{T},$ are adequate to distinguish regular and chaotic nature of the motion. However, one should keep in mind that these measures are geometric characteristics rather than dynamical invariants of the dynamics. Therefore, the above mentioned contrast between quasi-periodic and chaotic behavior can only be made by appropriate hypothesis tests.  

Further advantage of the present idea is that it uses directly the measured data set and requires neither Monte-Carlo simulation to achieve the best-fitting planetary model nor costly n-body integration. Consequently, the operation needs substantially shorter time to achieve the result. For example, the analysis of a time series with 950 data points in addition with the 100 surrogates requires no more than 10 minutes on a medium desktop machine. 

We emphasize that 950 measurements for both signals (RV, TTV) seem to be a sufficient amount of data to restore the dynamical operation given by the same length of numerical integration. Extending the data set with the length of the integration time makes change to the embedding parameters, nevertheless, the recurrence network analysis provides the same results for longer signals. It has to be noted that this method depends on the measured time series which is the past of the planetary dynamics and has a limited length evidently shorter than those obtained from numerical integration. Therefore, the conclusions drawn from the analysis should be treated in place and interpreted correctly, especially taking into account the time scales.

Moreover, the future surveys will serve further measurements to the existing ones generating longer time series that are going to be the basis of more trusty dynamical analysis. The proposed scheme can be generalized to more than two planets in a system \footnote{T. Kov\'acs, MNRAS (2019) in preparation}, thus, it can be used as a supportive method to the current efficient stability investigations.

\begin{acknowledgments}
Special thank is devoted to Tam\'as T\'el his valuable comments. The author is also grateful to the anonymous referee(s) for important remarks.
This work was supported by the NKFIH Hungarian Grants K119993, PD121223. The financial support of Bolyai Research Fellowship is also acknowledged.
\end{acknowledgments}

\bibliography{RecNet}

\begin{thebibliography}{36}%
\makeatletter
\providecommand \@ifxundefined [1]{%
 \@ifx{#1\undefined}
}%
\providecommand \@ifnum [1]{%
 \ifnum #1\expandafter \@firstoftwo
 \else \expandafter \@secondoftwo
 \fi
}%
\providecommand \@ifx [1]{%
 \ifx #1\expandafter \@firstoftwo
 \else \expandafter \@secondoftwo
 \fi
}%
\providecommand \natexlab [1]{#1}%
\providecommand \enquote  [1]{``#1''}%
\providecommand \bibnamefont  [1]{#1}%
\providecommand \bibfnamefont [1]{#1}%
\providecommand \citenamefont [1]{#1}%
\providecommand \href@noop [0]{\@secondoftwo}%
\providecommand \href [0]{\begingroup \@sanitize@url \@href}%
\providecommand \@href[1]{\@@startlink{#1}\@@href}%
\providecommand \@@href[1]{\endgroup#1\@@endlink}%
\providecommand \@sanitize@url [0]{\catcode `\\12\catcode `\$12\catcode
  `\&12\catcode `\#12\catcode `\^12\catcode `\_12\catcode `\%12\relax}%
\providecommand \@@startlink[1]{}%
\providecommand \@@endlink[0]{}%
\providecommand \url  [0]{\begingroup\@sanitize@url \@url }%
\providecommand \@url [1]{\endgroup\@href {#1}{\urlprefix }}%
\providecommand \urlprefix  [0]{URL }%
\providecommand \Eprint [0]{\href }%
\providecommand \doibase [0]{http://dx.doi.org/}%
\providecommand \selectlanguage [0]{\@gobble}%
\providecommand \bibinfo  [0]{\@secondoftwo}%
\providecommand \bibfield  [0]{\@secondoftwo}%
\providecommand \translation [1]{[#1]}%
\providecommand \BibitemOpen [0]{}%
\providecommand \bibitemStop [0]{}%
\providecommand \bibitemNoStop [0]{.\EOS\space}%
\providecommand \EOS [0]{\spacefactor3000\relax}%
\providecommand \BibitemShut  [1]{\csname bibitem#1\endcsname}%
\let\auto@bib@innerbib\@empty
\bibitem [{\citenamefont {{Foreman-Mackey}}\ \emph {et~al.}(2013)\citenamefont
  {{Foreman-Mackey}}, \citenamefont {{Hogg}}, \citenamefont {{Lang}},\ and\
  \citenamefont {{Goodman}}}]{Foreman-Mackey2013}%
  \BibitemOpen
  \bibfield  {author} {\bibinfo {author} {\bibfnamefont {D.}~\bibnamefont
  {{Foreman-Mackey}}}, \bibinfo {author} {\bibfnamefont {D.~W.}\ \bibnamefont
  {{Hogg}}}, \bibinfo {author} {\bibfnamefont {D.}~\bibnamefont {{Lang}}}, \
  and\ \bibinfo {author} {\bibfnamefont {J.}~\bibnamefont {{Goodman}}},\
  }\bibfield  {title} {\enquote {\bibinfo {title} {{emcee: The MCMC Hammer}},}\
  }\href {\doibase 10.1086/670067} {\bibfield  {journal} {\bibinfo  {journal}
  {Publications of the ASP}\ }\textbf {\bibinfo {volume} {125}},\ \bibinfo
  {pages} {306} (\bibinfo {year} {2013})},\ \Eprint
  {http://arxiv.org/abs/1202.3665} {arXiv:1202.3665 [astro-ph.IM]} \BibitemShut
  {NoStop}%
\bibitem [{\citenamefont {{Fabrycky}}(2010)}]{Fabrycky2010}%
  \BibitemOpen
  \bibfield  {author} {\bibinfo {author} {\bibfnamefont {D.~C.}\ \bibnamefont
  {{Fabrycky}}},\ }\bibfield  {title} {\enquote {\bibinfo {title}
  {{Non-Keplerian Dynamics of Exoplanets}},}\ }in\ \href@noop {} {\emph
  {\bibinfo {booktitle} {Exoplanets}}},\ \bibinfo {editor} {edited by\ \bibinfo
  {editor} {\bibfnamefont {S.}~\bibnamefont {{Seager}}}}\ (\bibinfo {year}
  {2010})\ pp.\ \bibinfo {pages} {217--238}\BibitemShut {NoStop}%
\bibitem [{\citenamefont {{Cincotta}}\ and\ \citenamefont
  {{Sim{\'o}}}(2000)}]{Cincotta2000}%
  \BibitemOpen
  \bibfield  {author} {\bibinfo {author} {\bibfnamefont {P.~M.}\ \bibnamefont
  {{Cincotta}}}\ and\ \bibinfo {author} {\bibfnamefont {C.}~\bibnamefont
  {{Sim{\'o}}}},\ }\bibfield  {title} {\enquote {\bibinfo {title} {{Simple
  tools to study global dynamics in non-axisymmetric galactic potentials -
  I}},}\ }\href {\doibase 10.1051/aas:2000108} {\bibfield  {journal} {\bibinfo
  {journal} {Astronomy and Astrophysics, Supplement}\ }\textbf {\bibinfo
  {volume} {147}},\ \bibinfo {pages} {205--228} (\bibinfo {year}
  {2000})}\BibitemShut {NoStop}%
\bibitem [{\citenamefont {{Rivera}}\ and\ \citenamefont
  {{Lissauer}}(2001)}]{Rivera2001}%
  \BibitemOpen
  \bibfield  {author} {\bibinfo {author} {\bibfnamefont {E.~J.}\ \bibnamefont
  {{Rivera}}}\ and\ \bibinfo {author} {\bibfnamefont {J.~J.}\ \bibnamefont
  {{Lissauer}}},\ }\bibfield  {title} {\enquote {\bibinfo {title} {{Dynamical
  Models of the Resonant Pair of Planets Orbiting the Star GJ 876}},}\ }\href
  {\doibase 10.1086/322477} {\bibfield  {journal} {\bibinfo  {journal} {The
  Astrophysical Journal}\ }\textbf {\bibinfo {volume} {558}},\ \bibinfo {pages}
  {392--402} (\bibinfo {year} {2001})}\BibitemShut {NoStop}%
\bibitem [{\citenamefont {{Armstrong}}\ \emph {et~al.}(2015)\citenamefont
  {{Armstrong}}, \citenamefont {{Santerne}}, \citenamefont {{Veras}},
  \citenamefont {{Barros}}, \citenamefont {{Demangeon}}, \citenamefont
  {{Lillo-Box}}, \citenamefont {{McCormac}}, \citenamefont {{Osborn}},
  \citenamefont {{Tsantaki}}, \citenamefont {{Almenara}}, \citenamefont
  {{Barrado}}, \citenamefont {{Boisse}}, \citenamefont {{Bonomo}},
  \citenamefont {{Brown}}, \citenamefont {{Bruno}}, \citenamefont {{Rey
  Cerda}}, \citenamefont {{Courcol}}, \citenamefont {{Deleuil}}, \citenamefont
  {{D{\'{\i}}az}}, \citenamefont {{Doyle}}, \citenamefont {{H{\'e}brard}},
  \citenamefont {{Kirk}}, \citenamefont {{Lam}}, \citenamefont {{Pollacco}},
  \citenamefont {{Rajpurohit}}, \citenamefont {{Spake}},\ and\ \citenamefont
  {{Walker}}}]{Armstrong2015}%
  \BibitemOpen
  \bibfield  {author} {\bibinfo {author} {\bibfnamefont {D.~J.}\ \bibnamefont
  {{Armstrong}}}, \bibinfo {author} {\bibfnamefont {A.}~\bibnamefont
  {{Santerne}}}, \bibinfo {author} {\bibfnamefont {D.}~\bibnamefont {{Veras}}},
  \bibinfo {author} {\bibfnamefont {S.~C.~C.}\ \bibnamefont {{Barros}}},
  \bibinfo {author} {\bibfnamefont {O.}~\bibnamefont {{Demangeon}}}, \bibinfo
  {author} {\bibfnamefont {J.}~\bibnamefont {{Lillo-Box}}}, \bibinfo {author}
  {\bibfnamefont {J.}~\bibnamefont {{McCormac}}}, \bibinfo {author}
  {\bibfnamefont {H.~P.}\ \bibnamefont {{Osborn}}}, \bibinfo {author}
  {\bibfnamefont {M.}~\bibnamefont {{Tsantaki}}}, \bibinfo {author}
  {\bibfnamefont {J.-M.}\ \bibnamefont {{Almenara}}}, \bibinfo {author}
  {\bibfnamefont {D.}~\bibnamefont {{Barrado}}}, \bibinfo {author}
  {\bibfnamefont {I.}~\bibnamefont {{Boisse}}}, \bibinfo {author}
  {\bibfnamefont {A.~S.}\ \bibnamefont {{Bonomo}}}, \bibinfo {author}
  {\bibfnamefont {D.~J.~A.}\ \bibnamefont {{Brown}}}, \bibinfo {author}
  {\bibfnamefont {G.}~\bibnamefont {{Bruno}}}, \bibinfo {author} {\bibfnamefont
  {J.}~\bibnamefont {{Rey Cerda}}}, \bibinfo {author} {\bibfnamefont
  {B.}~\bibnamefont {{Courcol}}}, \bibinfo {author} {\bibfnamefont
  {M.}~\bibnamefont {{Deleuil}}}, \bibinfo {author} {\bibfnamefont {R.~F.}\
  \bibnamefont {{D{\'{\i}}az}}}, \bibinfo {author} {\bibfnamefont {A.~P.}\
  \bibnamefont {{Doyle}}}, \bibinfo {author} {\bibfnamefont {G.}~\bibnamefont
  {{H{\'e}brard}}}, \bibinfo {author} {\bibfnamefont {J.}~\bibnamefont
  {{Kirk}}}, \bibinfo {author} {\bibfnamefont {K.~W.~F.}\ \bibnamefont
  {{Lam}}}, \bibinfo {author} {\bibfnamefont {D.~L.}\ \bibnamefont
  {{Pollacco}}}, \bibinfo {author} {\bibfnamefont {A.}~\bibnamefont
  {{Rajpurohit}}}, \bibinfo {author} {\bibfnamefont {J.}~\bibnamefont
  {{Spake}}}, \ and\ \bibinfo {author} {\bibfnamefont {S.~R.}\ \bibnamefont
  {{Walker}}},\ }\bibfield  {title} {\enquote {\bibinfo {title} {{One of the
  closest exoplanet pairs to the 3:2 mean motion resonance: K2-19b and c}},}\
  }\href {\doibase 10.1051/0004-6361/201526008} {\bibfield  {journal} {\bibinfo
   {journal} {Astronomy and Astrophysics}\ }\textbf {\bibinfo {volume} {582}},\
  \bibinfo {eid} {A33} (\bibinfo {year} {2015})},\ \Eprint
  {http://arxiv.org/abs/1503.00692} {arXiv:1503.00692 [astro-ph.EP]}
  \BibitemShut {NoStop}%
\bibitem [{\citenamefont {{Batygin}}, \citenamefont {{Deck}},\ and\
  \citenamefont {{Holman}}(2015)}]{Batygin2015}%
  \BibitemOpen
  \bibfield  {author} {\bibinfo {author} {\bibfnamefont {K.}~\bibnamefont
  {{Batygin}}}, \bibinfo {author} {\bibfnamefont {K.~M.}\ \bibnamefont
  {{Deck}}}, \ and\ \bibinfo {author} {\bibfnamefont {M.~J.}\ \bibnamefont
  {{Holman}}},\ }\bibfield  {title} {\enquote {\bibinfo {title} {{Dynamical
  Evolution of Multi-resonant Systems: The Case of GJ876}},}\ }\href {\doibase
  10.1088/0004-6256/149/5/167} {\bibfield  {journal} {\bibinfo  {journal} {The
  Astronomical Journal}\ }\textbf {\bibinfo {volume} {149}},\ \bibinfo {eid}
  {167} (\bibinfo {year} {2015})},\ \Eprint {http://arxiv.org/abs/1504.00051}
  {arXiv:1504.00051 [astro-ph.EP]} \BibitemShut {NoStop}%
\bibitem [{\citenamefont {Deck}\ \emph {et~al.}(2012)\citenamefont {Deck},
  \citenamefont {Holman}, \citenamefont {Agol}, \citenamefont {Carter},
  \citenamefont {Lissauer}, \citenamefont {Ragozzine},\ and\ \citenamefont
  {Winn}}]{Deck2012}%
  \BibitemOpen
  \bibfield  {author} {\bibinfo {author} {\bibfnamefont {K.~M.}\ \bibnamefont
  {Deck}}, \bibinfo {author} {\bibfnamefont {M.~J.}\ \bibnamefont {Holman}},
  \bibinfo {author} {\bibfnamefont {E.}~\bibnamefont {Agol}}, \bibinfo {author}
  {\bibfnamefont {J.~A.}\ \bibnamefont {Carter}}, \bibinfo {author}
  {\bibfnamefont {J.~J.}\ \bibnamefont {Lissauer}}, \bibinfo {author}
  {\bibfnamefont {D.}~\bibnamefont {Ragozzine}}, \ and\ \bibinfo {author}
  {\bibfnamefont {J.~N.}\ \bibnamefont {Winn}},\ }\bibfield  {title} {\enquote
  {\bibinfo {title} {{RAPID DYNAMICAL CHAOS IN AN EXOPLANETARY SYSTEM}},}\
  }\href@noop {} {\bibfield  {journal} {\bibinfo  {journal} {The Astrophysical
  Journall}\ }\textbf {\bibinfo {volume} {755}},\ \bibinfo {pages} {21}
  (\bibinfo {year} {2012})}\BibitemShut {NoStop}%
\bibitem [{\citenamefont {{Panichi}}, \citenamefont {{Go{\'z}dziewski}},\ and\
  \citenamefont {{Turchetti}}(2017)}]{Panichi2017}%
  \BibitemOpen
  \bibfield  {author} {\bibinfo {author} {\bibfnamefont {F.}~\bibnamefont
  {{Panichi}}}, \bibinfo {author} {\bibfnamefont {K.}~\bibnamefont
  {{Go{\'z}dziewski}}}, \ and\ \bibinfo {author} {\bibfnamefont
  {G.}~\bibnamefont {{Turchetti}}},\ }\bibfield  {title} {\enquote {\bibinfo
  {title} {{The reversibility error method (REM): a new, dynamical fast
  indicator for planetary dynamics}},}\ }\href {\doibase 10.1093/mnras/stx374}
  {\bibfield  {journal} {\bibinfo  {journal} {Mothly Notices of the Royal
  Astronomical Society}\ }\textbf {\bibinfo {volume} {468}},\ \bibinfo {pages}
  {469--491} (\bibinfo {year} {2017})}\BibitemShut {NoStop}%
\bibitem [{Note1()}]{Note1}%
  \BibitemOpen
  \bibinfo {note} {As a consequence of the mutual gravitational interaction of
  the planets one might observe slight deviations in orbital periods called
  transit timing variation.}\BibitemShut {Stop}%
\bibitem [{\citenamefont {{Kantz}}\ and\ \citenamefont
  {{Schreiber}}(2003)}]{Kantz2003}%
  \BibitemOpen
  \bibfield  {author} {\bibinfo {author} {\bibfnamefont {H.}~\bibnamefont
  {{Kantz}}}\ and\ \bibinfo {author} {\bibfnamefont {T.}~\bibnamefont
  {{Schreiber}}},\ }\href@noop {} {\emph {\bibinfo {title} {Nonlinear Time
  Series Analysis, by Holger Kantz , Thomas Schreiber, Cambridge, UK: Cambridge
  University Press, 2003}}}\ (\bibinfo  {publisher} {Cambridge University
  Press},\ \bibinfo {year} {2003})\BibitemShut {NoStop}%
\bibitem [{\citenamefont {{Donner}}\ \emph {et~al.}(2010)\citenamefont
  {{Donner}}, \citenamefont {{Zou}}, \citenamefont {{Donges}}, \citenamefont
  {{Marwan}},\ and\ \citenamefont {{Kurths}}}]{Donner2010}%
  \BibitemOpen
  \bibfield  {author} {\bibinfo {author} {\bibfnamefont {R.~V.}\ \bibnamefont
  {{Donner}}}, \bibinfo {author} {\bibfnamefont {Y.}~\bibnamefont {{Zou}}},
  \bibinfo {author} {\bibfnamefont {J.~F.}\ \bibnamefont {{Donges}}}, \bibinfo
  {author} {\bibfnamefont {N.}~\bibnamefont {{Marwan}}}, \ and\ \bibinfo
  {author} {\bibfnamefont {J.}~\bibnamefont {{Kurths}}},\ }\bibfield  {title}
  {\enquote {\bibinfo {title} {{Recurrence networks a novel paradigm for
  nonlinear time series analysis}},}\ }\href {\doibase
  10.1088/1367-2630/12/3/033025} {\bibfield  {journal} {\bibinfo  {journal}
  {New Journal of Physics}\ }\textbf {\bibinfo {volume} {12}},\ \bibinfo {eid}
  {033025} (\bibinfo {year} {2010})}\BibitemShut {NoStop}%
\bibitem [{\citenamefont {{Takens}}(1981)}]{Takens1981}%
  \BibitemOpen
  \bibfield  {author} {\bibinfo {author} {\bibfnamefont {F.}~\bibnamefont
  {{Takens}}},\ }\bibfield  {title} {\enquote {\bibinfo {title} {{Detecting
  strange attractors in turbulence}},}\ }\href {\doibase 10.1007/BFb0091924}
  {\bibfield  {journal} {\bibinfo  {journal} {Lecture Notes in Mathematics,
  Berlin Springer Verlag}\ }\textbf {\bibinfo {volume} {898}},\ \bibinfo
  {pages} {366} (\bibinfo {year} {1981})}\BibitemShut {NoStop}%
\bibitem [{\citenamefont {Semmlow}\ and\ \citenamefont
  {Griffel}(2014)}]{Semmlow}%
  \BibitemOpen
  \bibfield  {author} {\bibinfo {author} {\bibfnamefont {J.~L.}\ \bibnamefont
  {Semmlow}}\ and\ \bibinfo {author} {\bibfnamefont {B.}~\bibnamefont
  {Griffel}},\ }\href@noop {} {\emph {\bibinfo {title} {{Biosignal and medical
  image processing}}}}\ (\bibinfo  {publisher} {CRC Press},\ \bibinfo {year}
  {2014})\ p.\ \bibinfo {pages} {608}\BibitemShut {NoStop}%
\bibitem [{\citenamefont {{Eckmann}}, \citenamefont {{Oliffson Kamphorst}},\
  and\ \citenamefont {{Ruelle}}(1987)}]{Eckmann1987}%
  \BibitemOpen
  \bibfield  {author} {\bibinfo {author} {\bibfnamefont {J.-P.}\ \bibnamefont
  {{Eckmann}}}, \bibinfo {author} {\bibfnamefont {S.}~\bibnamefont {{Oliffson
  Kamphorst}}}, \ and\ \bibinfo {author} {\bibfnamefont {D.}~\bibnamefont
  {{Ruelle}}},\ }\bibfield  {title} {\enquote {\bibinfo {title} {{Recurrence
  plots of dynamical systems}},}\ }\href {\doibase 10.1209/0295-5075/4/9/004}
  {\bibfield  {journal} {\bibinfo  {journal} {EPL (Europhysics Letters)}\
  }\textbf {\bibinfo {volume} {4}},\ \bibinfo {pages} {973} (\bibinfo {year}
  {1987})}\BibitemShut {NoStop}%
\bibitem [{\citenamefont {{Zbilut}}\ and\ \citenamefont
  {{Webber}}(1992)}]{Zbilut1992}%
  \BibitemOpen
  \bibfield  {author} {\bibinfo {author} {\bibfnamefont {J.~P.}\ \bibnamefont
  {{Zbilut}}}\ and\ \bibinfo {author} {\bibfnamefont {C.~L.}\ \bibnamefont
  {{Webber}}},\ }\bibfield  {title} {\enquote {\bibinfo {title} {{Embeddings
  and delays as derived from quantification of recurrence plots}},}\ }\href
  {\doibase 10.1016/0375-9601(92)90426-M} {\bibfield  {journal} {\bibinfo
  {journal} {Physics Letters A}\ }\textbf {\bibinfo {volume} {171}},\ \bibinfo
  {pages} {199--203} (\bibinfo {year} {1992})}\BibitemShut {NoStop}%
\bibitem [{\citenamefont {Webber}\ and\ \citenamefont
  {Zbilut}(1994)}]{Webber1994}%
  \BibitemOpen
  \bibfield  {author} {\bibinfo {author} {\bibfnamefont {C.~L.}\ \bibnamefont
  {Webber}}\ and\ \bibinfo {author} {\bibfnamefont {J.~P.}\ \bibnamefont
  {Zbilut}},\ }\bibfield  {title} {\enquote {\bibinfo {title} {{Dynamical
  assessment of physiological systems and states using recurrence plot
  strategies.}}}\ }\href@noop {} {\bibfield  {journal} {\bibinfo  {journal}
  {Journal of applied physiology}\ }\textbf {\bibinfo {volume} {76 2}},\
  \bibinfo {pages} {965--73} (\bibinfo {year} {1994})}\BibitemShut {NoStop}%
\bibitem [{Note2()}]{Note2}%
  \BibitemOpen
  \bibinfo {note} {The visual interpretation of the matrix $\protect \mathbf
  {R}$ can achieved by plotting a dot when the matrix element is 1 and leave it
  empty otherwise.}\BibitemShut {Stop}%
\bibitem [{\citenamefont {Donner}\ \emph {et~al.}(2011)\citenamefont {Donner},
  \citenamefont {Heitzig}, \citenamefont {Donges}, \citenamefont {Zou},
  \citenamefont {Marwan},\ and\ \citenamefont {Kurths}}]{Donner2011}%
  \BibitemOpen
  \bibfield  {author} {\bibinfo {author} {\bibfnamefont {R.~V.}\ \bibnamefont
  {Donner}}, \bibinfo {author} {\bibfnamefont {J.}~\bibnamefont {Heitzig}},
  \bibinfo {author} {\bibfnamefont {J.~F.}\ \bibnamefont {Donges}}, \bibinfo
  {author} {\bibfnamefont {Y.}~\bibnamefont {Zou}}, \bibinfo {author}
  {\bibfnamefont {N.}~\bibnamefont {Marwan}}, \ and\ \bibinfo {author}
  {\bibfnamefont {J.}~\bibnamefont {Kurths}},\ }\bibfield  {title} {\enquote
  {\bibinfo {title} {{The geometry of chaotic dynamics – a complex network
  perspective}},}\ }\href {\doibase 10.1140/epjb/e2011-10899-1} {\bibfield
  {journal} {\bibinfo  {journal} {Eur. Phys. J. B}\ }\textbf {\bibinfo {volume}
  {84}},\ \bibinfo {pages} {653--672} (\bibinfo {year} {2011})}\BibitemShut
  {NoStop}%
\bibitem [{\citenamefont {{Zou}}\ \emph {et~al.}(2019)\citenamefont {{Zou}},
  \citenamefont {{Donner}}, \citenamefont {{Marwan}}, \citenamefont
  {{Donges}},\ and\ \citenamefont {{Kurths}}}]{Zou2019}%
  \BibitemOpen
  \bibfield  {author} {\bibinfo {author} {\bibfnamefont {Y.}~\bibnamefont
  {{Zou}}}, \bibinfo {author} {\bibfnamefont {R.~V.}\ \bibnamefont {{Donner}}},
  \bibinfo {author} {\bibfnamefont {N.}~\bibnamefont {{Marwan}}}, \bibinfo
  {author} {\bibfnamefont {J.~F.}\ \bibnamefont {{Donges}}}, \ and\ \bibinfo
  {author} {\bibfnamefont {J.}~\bibnamefont {{Kurths}}},\ }\bibfield  {title}
  {\enquote {\bibinfo {title} {{Complex network approaches to nonlinear time
  series analysis}},}\ }\href {\doibase 10.1016/j.physrep.2018.10.005}
  {\bibfield  {journal} {\bibinfo  {journal} {Physics Reports}\ }\textbf
  {\bibinfo {volume} {787}},\ \bibinfo {pages} {1--97} (\bibinfo {year}
  {2019})}\BibitemShut {NoStop}%
\bibitem [{\citenamefont {{Rein}}\ and\ \citenamefont
  {{Tamayo}}(2015)}]{Rein2015}%
  \BibitemOpen
  \bibfield  {author} {\bibinfo {author} {\bibfnamefont {H.}~\bibnamefont
  {{Rein}}}\ and\ \bibinfo {author} {\bibfnamefont {D.}~\bibnamefont
  {{Tamayo}}},\ }\bibfield  {title} {\enquote {\bibinfo {title} {{WHFAST: a
  fast and unbiased implementation of a symplectic Wisdom-Holman integrator for
  long-term gravitational simulations}},}\ }\href {\doibase
  10.1093/mnras/stv1257} {\bibfield  {journal} {\bibinfo  {journal} {Mothly
  Notices of the Royal Astronomical Society}\ }\textbf {\bibinfo {volume}
  {452}},\ \bibinfo {pages} {376--388} (\bibinfo {year} {2015})}\BibitemShut
  {NoStop}%
\bibitem [{\citenamefont {{Gillon}}\ \emph {et~al.}(2017)\citenamefont
  {{Gillon}}, \citenamefont {{Triaud}}, \citenamefont {{Demory}}, \citenamefont
  {{Jehin}}, \citenamefont {{Agol}}, \citenamefont {{Deck}}, \citenamefont
  {{Lederer}}, \citenamefont {{de Wit}}, \citenamefont {{Burdanov}},
  \citenamefont {{Ingalls}}, \citenamefont {{Bolmont}}, \citenamefont
  {{Leconte}}, \citenamefont {{Raymond}}, \citenamefont {{Selsis}},
  \citenamefont {{Turbet}}, \citenamefont {{Barkaoui}}, \citenamefont
  {{Burgasser}}, \citenamefont {{Burleigh}}, \citenamefont {{Carey}},
  \citenamefont {{Chaushev}}, \citenamefont {{Copperwheat}}, \citenamefont
  {{Delrez}}, \citenamefont {{Fernandes}}, \citenamefont {{Holdsworth}},
  \citenamefont {{Kotze}}, \citenamefont {{Van Grootel}}, \citenamefont
  {{Almleaky}}, \citenamefont {{Benkhaldoun}}, \citenamefont {{Magain}},\ and\
  \citenamefont {{Queloz}}}]{Gillon2017}%
  \BibitemOpen
  \bibfield  {author} {\bibinfo {author} {\bibfnamefont {M.}~\bibnamefont
  {{Gillon}}}, \bibinfo {author} {\bibfnamefont {A.~H.~M.~J.}\ \bibnamefont
  {{Triaud}}}, \bibinfo {author} {\bibfnamefont {B.-O.}\ \bibnamefont
  {{Demory}}}, \bibinfo {author} {\bibfnamefont {E.}~\bibnamefont {{Jehin}}},
  \bibinfo {author} {\bibfnamefont {E.}~\bibnamefont {{Agol}}}, \bibinfo
  {author} {\bibfnamefont {K.~M.}\ \bibnamefont {{Deck}}}, \bibinfo {author}
  {\bibfnamefont {S.~M.}\ \bibnamefont {{Lederer}}}, \bibinfo {author}
  {\bibfnamefont {J.}~\bibnamefont {{de Wit}}}, \bibinfo {author}
  {\bibfnamefont {A.}~\bibnamefont {{Burdanov}}}, \bibinfo {author}
  {\bibfnamefont {J.~G.}\ \bibnamefont {{Ingalls}}}, \bibinfo {author}
  {\bibfnamefont {E.}~\bibnamefont {{Bolmont}}}, \bibinfo {author}
  {\bibfnamefont {J.}~\bibnamefont {{Leconte}}}, \bibinfo {author}
  {\bibfnamefont {S.~N.}\ \bibnamefont {{Raymond}}}, \bibinfo {author}
  {\bibfnamefont {F.}~\bibnamefont {{Selsis}}}, \bibinfo {author}
  {\bibfnamefont {M.}~\bibnamefont {{Turbet}}}, \bibinfo {author}
  {\bibfnamefont {K.}~\bibnamefont {{Barkaoui}}}, \bibinfo {author}
  {\bibfnamefont {A.}~\bibnamefont {{Burgasser}}}, \bibinfo {author}
  {\bibfnamefont {M.~R.}\ \bibnamefont {{Burleigh}}}, \bibinfo {author}
  {\bibfnamefont {S.~J.}\ \bibnamefont {{Carey}}}, \bibinfo {author}
  {\bibfnamefont {A.}~\bibnamefont {{Chaushev}}}, \bibinfo {author}
  {\bibfnamefont {C.~M.}\ \bibnamefont {{Copperwheat}}}, \bibinfo {author}
  {\bibfnamefont {L.}~\bibnamefont {{Delrez}}}, \bibinfo {author}
  {\bibfnamefont {C.~S.}\ \bibnamefont {{Fernandes}}}, \bibinfo {author}
  {\bibfnamefont {D.~L.}\ \bibnamefont {{Holdsworth}}}, \bibinfo {author}
  {\bibfnamefont {E.~J.}\ \bibnamefont {{Kotze}}}, \bibinfo {author}
  {\bibfnamefont {V.}~\bibnamefont {{Van Grootel}}}, \bibinfo {author}
  {\bibfnamefont {Y.}~\bibnamefont {{Almleaky}}}, \bibinfo {author}
  {\bibfnamefont {Z.}~\bibnamefont {{Benkhaldoun}}}, \bibinfo {author}
  {\bibfnamefont {P.}~\bibnamefont {{Magain}}}, \ and\ \bibinfo {author}
  {\bibfnamefont {D.}~\bibnamefont {{Queloz}}},\ }\bibfield  {title} {\enquote
  {\bibinfo {title} {{Seven temperate terrestrial planets around the nearby
  ultracool dwarf star TRAPPIST-1}},}\ }\href {\doibase 10.1038/nature21360}
  {\bibfield  {journal} {\bibinfo  {journal} {Nature}\ }\textbf {\bibinfo
  {volume} {542}},\ \bibinfo {pages} {456--460} (\bibinfo {year} {2017})},\
  \Eprint {http://arxiv.org/abs/1703.01424} {arXiv:1703.01424 [astro-ph.EP]}
  \BibitemShut {NoStop}%
\bibitem [{\citenamefont {{Deleuil}}\ \emph {et~al.}(2014)\citenamefont
  {{Deleuil}}, \citenamefont {{Almenara}}, \citenamefont {{Santerne}},
  \citenamefont {{Barros}}, \citenamefont {{Havel}}, \citenamefont
  {{H{\'e}brard}}, \citenamefont {{Bonomo}}, \citenamefont {{Bouchy}},
  \citenamefont {{Bruno}}, \citenamefont {{Damiani}}, \citenamefont
  {{D{\'{\i}}az}}, \citenamefont {{Montagnier}},\ and\ \citenamefont
  {{Moutou}}}]{Deleuil2014}%
  \BibitemOpen
  \bibfield  {author} {\bibinfo {author} {\bibfnamefont {M.}~\bibnamefont
  {{Deleuil}}}, \bibinfo {author} {\bibfnamefont {J.-M.}\ \bibnamefont
  {{Almenara}}}, \bibinfo {author} {\bibfnamefont {A.}~\bibnamefont
  {{Santerne}}}, \bibinfo {author} {\bibfnamefont {S.~C.~C.}\ \bibnamefont
  {{Barros}}}, \bibinfo {author} {\bibfnamefont {M.}~\bibnamefont {{Havel}}},
  \bibinfo {author} {\bibfnamefont {G.}~\bibnamefont {{H{\'e}brard}}}, \bibinfo
  {author} {\bibfnamefont {A.~S.}\ \bibnamefont {{Bonomo}}}, \bibinfo {author}
  {\bibfnamefont {F.}~\bibnamefont {{Bouchy}}}, \bibinfo {author}
  {\bibfnamefont {G.}~\bibnamefont {{Bruno}}}, \bibinfo {author} {\bibfnamefont
  {C.}~\bibnamefont {{Damiani}}}, \bibinfo {author} {\bibfnamefont {R.~F.}\
  \bibnamefont {{D{\'{\i}}az}}}, \bibinfo {author} {\bibfnamefont
  {G.}~\bibnamefont {{Montagnier}}}, \ and\ \bibinfo {author} {\bibfnamefont
  {C.}~\bibnamefont {{Moutou}}},\ }\bibfield  {title} {\enquote {\bibinfo
  {title} {{SOPHIE velocimetry of Kepler transit candidates XI. Kepler-412
  system: probing the properties of a new inflated hot Jupiter}},}\ }\href
  {\doibase 10.1051/0004-6361/201323017} {\bibfield  {journal} {\bibinfo
  {journal} {Astronomy and Astrophysics}\ }\textbf {\bibinfo {volume} {564}},\
  \bibinfo {pages} {A56} (\bibinfo {year} {2014})},\ \Eprint
  {http://arxiv.org/abs/1401.6811} {arXiv:1401.6811 [astro-ph.EP]} \BibitemShut
  {NoStop}%
\bibitem [{\citenamefont {{Hegger}}, \citenamefont {{Kantz}},\ and\
  \citenamefont {{Schreiber}}(1999)}]{Hegger1999}%
  \BibitemOpen
  \bibfield  {author} {\bibinfo {author} {\bibfnamefont {R.}~\bibnamefont
  {{Hegger}}}, \bibinfo {author} {\bibfnamefont {H.}~\bibnamefont {{Kantz}}}, \
  and\ \bibinfo {author} {\bibfnamefont {T.}~\bibnamefont {{Schreiber}}},\
  }\bibfield  {title} {\enquote {\bibinfo {title} {{Practical implementation of
  nonlinear time series methods: The TISEAN package}},}\ }\href {\doibase
  10.1063/1.166424} {\bibfield  {journal} {\bibinfo  {journal} {Chaos}\
  }\textbf {\bibinfo {volume} {9}},\ \bibinfo {pages} {413--435} (\bibinfo
  {year} {1999})}\BibitemShut {NoStop}%
\bibitem [{\citenamefont {{Donges}}\ \emph {et~al.}(2015)\citenamefont
  {{Donges}}, \citenamefont {{Heitzig}}, \citenamefont {{Beronov}},
  \citenamefont {{Wiedermann}}, \citenamefont {{Runge}}, \citenamefont
  {{Feng}}, \citenamefont {{Tupikina}}, \citenamefont {{Stolbova}},
  \citenamefont {{Donner}}, \citenamefont {{Marwan}}, \citenamefont
  {{Dijkstra}},\ and\ \citenamefont {{Kurths}}}]{Donges2015}%
  \BibitemOpen
  \bibfield  {author} {\bibinfo {author} {\bibfnamefont {J.~F.}\ \bibnamefont
  {{Donges}}}, \bibinfo {author} {\bibfnamefont {J.}~\bibnamefont {{Heitzig}}},
  \bibinfo {author} {\bibfnamefont {B.}~\bibnamefont {{Beronov}}}, \bibinfo
  {author} {\bibfnamefont {M.}~\bibnamefont {{Wiedermann}}}, \bibinfo {author}
  {\bibfnamefont {J.}~\bibnamefont {{Runge}}}, \bibinfo {author} {\bibfnamefont
  {Q.~Y.}\ \bibnamefont {{Feng}}}, \bibinfo {author} {\bibfnamefont
  {L.}~\bibnamefont {{Tupikina}}}, \bibinfo {author} {\bibfnamefont
  {V.}~\bibnamefont {{Stolbova}}}, \bibinfo {author} {\bibfnamefont {R.~V.}\
  \bibnamefont {{Donner}}}, \bibinfo {author} {\bibfnamefont {N.}~\bibnamefont
  {{Marwan}}}, \bibinfo {author} {\bibfnamefont {H.~A.}\ \bibnamefont
  {{Dijkstra}}}, \ and\ \bibinfo {author} {\bibfnamefont {J.}~\bibnamefont
  {{Kurths}}},\ }\bibfield  {title} {\enquote {\bibinfo {title} {{Unified
  functional network and nonlinear time series analysis for complex systems
  science: The pyunicorn package}},}\ }\href {\doibase 10.1063/1.4934554}
  {\bibfield  {journal} {\bibinfo  {journal} {Chaos}\ }\textbf {\bibinfo
  {volume} {25}},\ \bibinfo {eid} {113101} (\bibinfo {year} {2015})},\ \Eprint
  {http://arxiv.org/abs/1507.01571} {arXiv:1507.01571 [physics.data-an]}
  \BibitemShut {NoStop}%
\bibitem [{\citenamefont {{Zou}}\ \emph {et~al.}(2016)\citenamefont {{Zou}},
  \citenamefont {{Donner}}, \citenamefont {{Thiel}},\ and\ \citenamefont
  {{Kurths}}}]{Zou2016}%
  \BibitemOpen
  \bibfield  {author} {\bibinfo {author} {\bibfnamefont {Y.}~\bibnamefont
  {{Zou}}}, \bibinfo {author} {\bibfnamefont {R.~V.}\ \bibnamefont {{Donner}}},
  \bibinfo {author} {\bibfnamefont {M.}~\bibnamefont {{Thiel}}}, \ and\
  \bibinfo {author} {\bibfnamefont {J.}~\bibnamefont {{Kurths}}},\ }\bibfield
  {title} {\enquote {\bibinfo {title} {{Disentangling regular and chaotic
  motion in the standard map using complex network analysis of recurrences in
  phase space}},}\ }\href {\doibase 10.1063/1.4942584} {\bibfield  {journal}
  {\bibinfo  {journal} {Chaos}\ }\textbf {\bibinfo {volume} {26}},\ \bibinfo
  {eid} {023120} (\bibinfo {year} {2016})}\BibitemShut {NoStop}%
\bibitem [{\citenamefont {{Contopoulos}}(2002)}]{Contopoulos2002}%
  \BibitemOpen
  \bibfield  {author} {\bibinfo {author} {\bibfnamefont {G.}~\bibnamefont
  {{Contopoulos}}},\ }\href@noop {} {\emph {\bibinfo {title} {Order and chaos
  in dynamical astronomy}}}\ (\bibinfo  {publisher} {Springer},\ \bibinfo
  {year} {2002})\BibitemShut {NoStop}%
\bibitem [{\citenamefont {{Zou}}\ \emph {et~al.}(2010)\citenamefont {{Zou}},
  \citenamefont {{Donner}}, \citenamefont {{Donges}}, \citenamefont
  {{Marwan}},\ and\ \citenamefont {{Kurths}}}]{Zou2010}%
  \BibitemOpen
  \bibfield  {author} {\bibinfo {author} {\bibfnamefont {Y.}~\bibnamefont
  {{Zou}}}, \bibinfo {author} {\bibfnamefont {R.~V.}\ \bibnamefont {{Donner}}},
  \bibinfo {author} {\bibfnamefont {J.~F.}\ \bibnamefont {{Donges}}}, \bibinfo
  {author} {\bibfnamefont {N.}~\bibnamefont {{Marwan}}}, \ and\ \bibinfo
  {author} {\bibfnamefont {J.}~\bibnamefont {{Kurths}}},\ }\bibfield  {title}
  {\enquote {\bibinfo {title} {{Identifying complex periodic windows in
  continuous-time dynamical systems using recurrence-based methods}},}\ }\href
  {\doibase 10.1063/1.3523304} {\bibfield  {journal} {\bibinfo  {journal}
  {Chaos}\ }\textbf {\bibinfo {volume} {20}},\ \bibinfo {eid} {043130}
  (\bibinfo {year} {2010})},\ \Eprint {http://arxiv.org/abs/1011.5172}
  {arXiv:1011.5172 [nlin.CD]} \BibitemShut {NoStop}%
\bibitem [{\citenamefont {Schreiber}\ and\ \citenamefont
  {Schmitz}(2000)}]{Schreiber2000}%
  \BibitemOpen
  \bibfield  {author} {\bibinfo {author} {\bibfnamefont {T.}~\bibnamefont
  {Schreiber}}\ and\ \bibinfo {author} {\bibfnamefont {A.}~\bibnamefont
  {Schmitz}},\ }\bibfield  {title} {\enquote {\bibinfo {title} {{Surrogate time
  series}},}\ }\href@noop {} {\bibfield  {journal} {\bibinfo  {journal}
  {Physica D: Nonlinear Phenomena}\ }\textbf {\bibinfo {volume} {142}},\
  \bibinfo {pages} {346--382} (\bibinfo {year} {2000})}\BibitemShut {NoStop}%
\bibitem [{\citenamefont {{Luo}}, \citenamefont {{Nakamura}},\ and\
  \citenamefont {{Small}}(2005)}]{Luo2005}%
  \BibitemOpen
  \bibfield  {author} {\bibinfo {author} {\bibfnamefont {X.}~\bibnamefont
  {{Luo}}}, \bibinfo {author} {\bibfnamefont {T.}~\bibnamefont {{Nakamura}}}, \
  and\ \bibinfo {author} {\bibfnamefont {M.}~\bibnamefont {{Small}}},\
  }\bibfield  {title} {\enquote {\bibinfo {title} {{Surrogate test to
  distinguish between chaotic and pseudoperiodic time series}},}\ }\href
  {\doibase 10.1103/PhysRevE.71.026230} {\bibfield  {journal} {\bibinfo
  {journal} {Physical Review E}\ }\textbf {\bibinfo {volume} {71}},\ \bibinfo
  {eid} {026230} (\bibinfo {year} {2005})},\ \Eprint
  {http://arxiv.org/abs/nlin/0404054} {nlin/0404054} \BibitemShut {NoStop}%
\bibitem [{\citenamefont {{Carri{\'o}n}}\ and\ \citenamefont
  {{Miralles}}(2016)}]{Carrion2016}%
  \BibitemOpen
  \bibfield  {author} {\bibinfo {author} {\bibfnamefont {A.}~\bibnamefont
  {{Carri{\'o}n}}}\ and\ \bibinfo {author} {\bibfnamefont {R.}~\bibnamefont
  {{Miralles}}},\ }\bibfield  {title} {\enquote {\bibinfo {title} {{New
  Insights for Testing Linearity and Complexity with Surrogates: A Recurrence
  Plot Approach}},}\ }in\ \href {\doibase 10.1007/978-3-319-29922-8_5} {\emph
  {\bibinfo {booktitle} {Recurrence Plots and Their Quantifications: Expanding
  Horizon: Proceedings of the 6th International Symposium on Recurrence
  Plots}}},\ \bibinfo {series} {Springer Proceedings in Physics}, Vol.\
  \bibinfo {volume} {180},\ \bibinfo {editor} {edited by\ \bibinfo {editor}
  {\bibfnamefont {C.~L.}\ \bibnamefont {{Webber}}, \bibfnamefont {Jr.}},
  \bibinfo {editor} {\bibfnamefont {C.}~\bibnamefont {{Ioana}}}, \ and\
  \bibinfo {editor} {\bibfnamefont {N.}~\bibnamefont {{Marwan}}}}\ (\bibinfo
  {year} {2016})\ p.~\bibinfo {pages} {91}\BibitemShut {NoStop}%
\bibitem [{\citenamefont {{Small}}, \citenamefont {{Yu}},\ and\ \citenamefont
  {{Harrison}}(2001)}]{Small2001}%
  \BibitemOpen
  \bibfield  {author} {\bibinfo {author} {\bibfnamefont {M.}~\bibnamefont
  {{Small}}}, \bibinfo {author} {\bibfnamefont {D.}~\bibnamefont {{Yu}}}, \
  and\ \bibinfo {author} {\bibfnamefont {R.~G.}\ \bibnamefont {{Harrison}}},\
  }\bibfield  {title} {\enquote {\bibinfo {title} {{Surrogate Test for
  Pseudoperiodic Time Series Data}},}\ }\href {\doibase
  10.1103/PhysRevLett.87.188101} {\bibfield  {journal} {\bibinfo  {journal}
  {Physical Review Letters}\ }\textbf {\bibinfo {volume} {87}},\ \bibinfo {eid}
  {188101} (\bibinfo {year} {2001})}\BibitemShut {NoStop}%
\bibitem [{\citenamefont {Thiel}\ \emph {et~al.}(2006)\citenamefont {Thiel},
  \citenamefont {Romano}, \citenamefont {Kurths}, \citenamefont {Rolfs},\ and\
  \citenamefont {Kliegl}}]{Thiel2006}%
  \BibitemOpen
  \bibfield  {author} {\bibinfo {author} {\bibfnamefont {M.}~\bibnamefont
  {Thiel}}, \bibinfo {author} {\bibfnamefont {M.~C.}\ \bibnamefont {Romano}},
  \bibinfo {author} {\bibfnamefont {J.}~\bibnamefont {Kurths}}, \bibinfo
  {author} {\bibfnamefont {M.}~\bibnamefont {Rolfs}}, \ and\ \bibinfo {author}
  {\bibfnamefont {R.}~\bibnamefont {Kliegl}},\ }\bibfield  {title} {\enquote
  {\bibinfo {title} {{Twin surrogates to test for complex synchronisation}},}\
  }\href {\doibase 10.1209/epl/i2006-10147-0} {\bibfield  {journal} {\bibinfo
  {journal} {Europhysics Letters}\ }\textbf {\bibinfo {volume} {75}},\ \bibinfo
  {pages} {535--541} (\bibinfo {year} {2006})}\BibitemShut {NoStop}%
\bibitem [{\citenamefont {{Lekscha}}\ and\ \citenamefont
  {{Donner}}(2018)}]{Lekscha2018}%
  \BibitemOpen
  \bibfield  {author} {\bibinfo {author} {\bibfnamefont {J.}~\bibnamefont
  {{Lekscha}}}\ and\ \bibinfo {author} {\bibfnamefont {R.~V.}\ \bibnamefont
  {{Donner}}},\ }\bibfield  {title} {\enquote {\bibinfo {title} {{Phase space
  reconstruction for non-uniformly sampled noisy time series}},}\ }\href
  {\doibase 10.1063/1.5023860} {\bibfield  {journal} {\bibinfo  {journal}
  {Chaos}\ }\textbf {\bibinfo {volume} {28}},\ \bibinfo {eid} {085702}
  (\bibinfo {year} {2018})},\ \Eprint {http://arxiv.org/abs/1801.09517}
  {arXiv:1801.09517 [physics.data-an]} \BibitemShut {NoStop}%
\bibitem [{\citenamefont {{Holczer}}\ \emph {et~al.}(2016)\citenamefont
  {{Holczer}}, \citenamefont {{Mazeh}}, \citenamefont {{Nachmani}},
  \citenamefont {{Jontof-Hutter}}, \citenamefont {{Ford}}, \citenamefont
  {{Fabrycky}}, \citenamefont {{Ragozzine}}, \citenamefont {{Kane}},\ and\
  \citenamefont {{Steffen}}}]{Holczer2016}%
  \BibitemOpen
  \bibfield  {author} {\bibinfo {author} {\bibfnamefont {T.}~\bibnamefont
  {{Holczer}}}, \bibinfo {author} {\bibfnamefont {T.}~\bibnamefont {{Mazeh}}},
  \bibinfo {author} {\bibfnamefont {G.}~\bibnamefont {{Nachmani}}}, \bibinfo
  {author} {\bibfnamefont {D.}~\bibnamefont {{Jontof-Hutter}}}, \bibinfo
  {author} {\bibfnamefont {E.~B.}\ \bibnamefont {{Ford}}}, \bibinfo {author}
  {\bibfnamefont {D.}~\bibnamefont {{Fabrycky}}}, \bibinfo {author}
  {\bibfnamefont {D.}~\bibnamefont {{Ragozzine}}}, \bibinfo {author}
  {\bibfnamefont {M.}~\bibnamefont {{Kane}}}, \ and\ \bibinfo {author}
  {\bibfnamefont {J.~H.}\ \bibnamefont {{Steffen}}},\ }\bibfield  {title}
  {\enquote {\bibinfo {title} {{Transit Timing Observations from Kepler. IX.
  Catalog of the Full Long-cadence Data Set}},}\ }\href {\doibase
  10.3847/0067-0049/225/1/9} {\bibfield  {journal} {\bibinfo  {journal} {The
  Astrophysical Journals}\ }\textbf {\bibinfo {volume} {225}},\ \bibinfo {eid}
  {9} (\bibinfo {year} {2016})},\ \Eprint {http://arxiv.org/abs/1606.01744}
  {arXiv:1606.01744 [astro-ph.EP]} \BibitemShut {NoStop}%
\bibitem [{\citenamefont {{Carter}}\ \emph {et~al.}(2012)\citenamefont
  {{Carter}}, \citenamefont {{Agol}}, \citenamefont {{Chaplin}}, \citenamefont
  {{Basu}}, \citenamefont {{Bedding}}, \citenamefont {{Buchhave}},
  \citenamefont {{Christensen-Dalsgaard}}, \citenamefont {{Deck}},
  \citenamefont {{Elsworth}}, \citenamefont {{Fabrycky}}, \citenamefont
  {{Ford}}, \citenamefont {{Fortney}}, \citenamefont {{Hale}}, \citenamefont
  {{Handberg}}, \citenamefont {{Hekker}}, \citenamefont {{Holman}},
  \citenamefont {{Huber}}, \citenamefont {{Karoff}}, \citenamefont {{Kawaler}},
  \citenamefont {{Kjeldsen}}, \citenamefont {{Lissauer}}, \citenamefont
  {{Lopez}}, \citenamefont {{Lund}}, \citenamefont {{Lundkvist}}, \citenamefont
  {{Metcalfe}}, \citenamefont {{Miglio}}, \citenamefont {{Rogers}},
  \citenamefont {{Stello}}, \citenamefont {{Borucki}}, \citenamefont
  {{Bryson}}, \citenamefont {{Christiansen}}, \citenamefont {{Cochran}},
  \citenamefont {{Geary}}, \citenamefont {{Gilliland}}, \citenamefont {{Haas}},
  \citenamefont {{Hall}}, \citenamefont {{Howard}}, \citenamefont {{Jenkins}},
  \citenamefont {{Klaus}}, \citenamefont {{Koch}}, \citenamefont {{Latham}},
  \citenamefont {{MacQueen}}, \citenamefont {{Sasselov}}, \citenamefont
  {{Steffen}}, \citenamefont {{Twicken}},\ and\ \citenamefont
  {{Winn}}}]{Carter2012}%
  \BibitemOpen
  \bibfield  {author} {\bibinfo {author} {\bibfnamefont {J.~A.}\ \bibnamefont
  {{Carter}}}, \bibinfo {author} {\bibfnamefont {E.}~\bibnamefont {{Agol}}},
  \bibinfo {author} {\bibfnamefont {W.~J.}\ \bibnamefont {{Chaplin}}}, \bibinfo
  {author} {\bibfnamefont {S.}~\bibnamefont {{Basu}}}, \bibinfo {author}
  {\bibfnamefont {T.~R.}\ \bibnamefont {{Bedding}}}, \bibinfo {author}
  {\bibfnamefont {L.~A.}\ \bibnamefont {{Buchhave}}}, \bibinfo {author}
  {\bibfnamefont {J.}~\bibnamefont {{Christensen-Dalsgaard}}}, \bibinfo
  {author} {\bibfnamefont {K.~M.}\ \bibnamefont {{Deck}}}, \bibinfo {author}
  {\bibfnamefont {Y.}~\bibnamefont {{Elsworth}}}, \bibinfo {author}
  {\bibfnamefont {D.~C.}\ \bibnamefont {{Fabrycky}}}, \bibinfo {author}
  {\bibfnamefont {E.~B.}\ \bibnamefont {{Ford}}}, \bibinfo {author}
  {\bibfnamefont {J.~J.}\ \bibnamefont {{Fortney}}}, \bibinfo {author}
  {\bibfnamefont {S.~J.}\ \bibnamefont {{Hale}}}, \bibinfo {author}
  {\bibfnamefont {R.}~\bibnamefont {{Handberg}}}, \bibinfo {author}
  {\bibfnamefont {S.}~\bibnamefont {{Hekker}}}, \bibinfo {author}
  {\bibfnamefont {M.~J.}\ \bibnamefont {{Holman}}}, \bibinfo {author}
  {\bibfnamefont {D.}~\bibnamefont {{Huber}}}, \bibinfo {author} {\bibfnamefont
  {C.}~\bibnamefont {{Karoff}}}, \bibinfo {author} {\bibfnamefont {S.~D.}\
  \bibnamefont {{Kawaler}}}, \bibinfo {author} {\bibfnamefont {H.}~\bibnamefont
  {{Kjeldsen}}}, \bibinfo {author} {\bibfnamefont {J.~J.}\ \bibnamefont
  {{Lissauer}}}, \bibinfo {author} {\bibfnamefont {E.~D.}\ \bibnamefont
  {{Lopez}}}, \bibinfo {author} {\bibfnamefont {M.~N.}\ \bibnamefont {{Lund}}},
  \bibinfo {author} {\bibfnamefont {M.}~\bibnamefont {{Lundkvist}}}, \bibinfo
  {author} {\bibfnamefont {T.~S.}\ \bibnamefont {{Metcalfe}}}, \bibinfo
  {author} {\bibfnamefont {A.}~\bibnamefont {{Miglio}}}, \bibinfo {author}
  {\bibfnamefont {L.~A.}\ \bibnamefont {{Rogers}}}, \bibinfo {author}
  {\bibfnamefont {D.}~\bibnamefont {{Stello}}}, \bibinfo {author}
  {\bibfnamefont {W.~J.}\ \bibnamefont {{Borucki}}}, \bibinfo {author}
  {\bibfnamefont {S.}~\bibnamefont {{Bryson}}}, \bibinfo {author}
  {\bibfnamefont {J.~L.}\ \bibnamefont {{Christiansen}}}, \bibinfo {author}
  {\bibfnamefont {W.~D.}\ \bibnamefont {{Cochran}}}, \bibinfo {author}
  {\bibfnamefont {J.~C.}\ \bibnamefont {{Geary}}}, \bibinfo {author}
  {\bibfnamefont {R.~L.}\ \bibnamefont {{Gilliland}}}, \bibinfo {author}
  {\bibfnamefont {M.~R.}\ \bibnamefont {{Haas}}}, \bibinfo {author}
  {\bibfnamefont {J.}~\bibnamefont {{Hall}}}, \bibinfo {author} {\bibfnamefont
  {A.~W.}\ \bibnamefont {{Howard}}}, \bibinfo {author} {\bibfnamefont {J.~M.}\
  \bibnamefont {{Jenkins}}}, \bibinfo {author} {\bibfnamefont {T.}~\bibnamefont
  {{Klaus}}}, \bibinfo {author} {\bibfnamefont {D.~G.}\ \bibnamefont {{Koch}}},
  \bibinfo {author} {\bibfnamefont {D.~W.}\ \bibnamefont {{Latham}}}, \bibinfo
  {author} {\bibfnamefont {P.~J.}\ \bibnamefont {{MacQueen}}}, \bibinfo
  {author} {\bibfnamefont {D.}~\bibnamefont {{Sasselov}}}, \bibinfo {author}
  {\bibfnamefont {J.~H.}\ \bibnamefont {{Steffen}}}, \bibinfo {author}
  {\bibfnamefont {J.~D.}\ \bibnamefont {{Twicken}}}, \ and\ \bibinfo {author}
  {\bibfnamefont {J.~N.}\ \bibnamefont {{Winn}}},\ }\bibfield  {title}
  {\enquote {\bibinfo {title} {{Kepler-36: A Pair of Planets with Neighboring
  Orbits and Dissimilar Densities}},}\ }\href {\doibase
  10.1126/science.1223269} {\bibfield  {journal} {\bibinfo  {journal}
  {Science}\ }\textbf {\bibinfo {volume} {337}},\ \bibinfo {pages} {556}
  (\bibinfo {year} {2012})},\ \Eprint {http://arxiv.org/abs/1206.4718}
  {arXiv:1206.4718 [astro-ph.EP]} \BibitemShut {NoStop}%
\bibitem [{Note3()}]{Note3}%
  \BibitemOpen
  \bibinfo {note} {T. Kov\'acs, MNRAS (2019) in preparation}\BibitemShut
  {NoStop}%
\end{thebibliography}%

\end{document}